\documentclass[11pt]{article}
\usepackage{epsfig}
\usepackage{amsfonts}
\usepackage{amssymb}
\usepackage{graphicx}
\usepackage{psfrag}
\usepackage{mathrsfs}
\usepackage{natbib}

\begin{document}
\def\bcolon{\mbox{\boldmath$\colon$}}
\def\bdot{\mbox{\boldmath$\cdot$}}
\def\bone{{\bf 1}}
\def\bzero{{\bf 0}}
\def\DIV {\hbox{\rm div}}
\def\Grad{\hbox{\rm Grad}}
\def\sym{\mathop{\rm sym}\nolimits}
\def\dev{\mathop{\rm dev}\nolimits}
\def\Dev{\mathop {\rm DEV}\nolimits}
\def\jmpdelu{{\lbrack\!\lbrack \Delta u\rbrack\!\rbrack}}
\def\jmpudot{{\lbrack\!\lbrack\dot u\rbrack\!\rbrack}}
\def\jmpu{{\lbrack\!\lbrack u\rbrack\!\rbrack}}
\def\jmphi{{\lbrack\!\lbrack\varphi\rbrack\!\rbrack}}
\def\ljmp{{\lbrack\!\lbrack}}
\def\rjmp{{\rbrack\!\rbrack}}
\def\sB{{\mathcal B}}
\def\sU{{\mathcal U}}
\def\sC{{\mathcal C}}
\def\sS{{\mathcal S}}
\def\sV{{\mathcal V}}
\def\sL{{\mathcal L}}
\def\sO{{\mathcal O}}
\def\sH{{\mathcal H}}
\def\sG{{\mathcal G}}
\def\sM{{\mathcal M}}
\def\sN{{\mathcal N}}
\def\sF{{\mathcal F}}
\def\sW{{\mathcal W}}
\def\sT{{\mathcal T}}
\def\sR{{\mathcal R}}
\def\sJ{{\mathcal J}}
\def\sK{{\mathcal K}}
\def\sE{{\mathcal E}}
\def\sS{{\mathcal S}}
\def\sH{{\mathcal H}}
\def\sD{{\mathcal D}}
\def\sG{{\mathcal G}}
\def\sP{{\mathcal P}}
\def\Bnabla{\mbox{\boldmath$\nabla$}}
\def\BGamma{\mbox{\boldmath$\Gamma$}}
\def\BDelta{\mbox{\boldmath$\Delta$}}
\def\BTheta{\mbox{\boldmath$\Theta$}}
\def\BLambda{\mbox{\boldmath$\Lambda$}}
\def\BXi{\mbox{\boldmath$\Xi$}}
\def\BPi{\mbox{\boldmath$\Pi$}}
\def\BSigma{\mbox{\boldmath$\Sigma$}}
\def\BUpsilon{\mbox{\boldmath$\Upsilon$}}
\def\BPhi{\mbox{\boldmath$\Phi$}}
\def\BPsi{\mbox{\boldmath$\Psi$}}
\def\BOmega{\mbox{\boldmath$\Omega$}}
\def\Balpha{\mbox{\boldmath$\alpha$}}
\def\Bbeta{\mbox{\boldmath$\beta$}}
\def\Bgamma{\mbox{\boldmath$\gamma$}}
\def\Bdelta{\mbox{\boldmath$\delta$}}
\def\Bepsilon{\mbox{\boldmath$\epsilon$}}
\def\Bzeta{\mbox{\boldmath$\zeta$}}
\def\Beta{\mbox{\boldmath$\eta$}}
\def\Btheta{\mbox{\boldmath$\theta$}}
\def\Biota{\mbox{\boldmath$\iota$}}
\def\Bkappa{\mbox{\boldmath$\kappa$}}
\def\Blambda{\mbox{\boldmath$\lambda$}}
\def\Bmu{\mbox{\boldmath$\mu$}}
\def\Bnu{\mbox{\boldmath$\nu$}}
\def\Bxi{\mbox{\boldmath$\xi$}}
\def\Bpi{\mbox{\boldmath$\pi$}}
\def\Brho{\mbox{\boldmath$\rho$}}
\def\Bsigma{\mbox{\boldmath$\sigma$}}
\def\Btau{\mbox{\boldmath$\tau$}}
\def\Bupsilon{\mbox{\boldmath$\upsilon$}}
\def\Bphi{\mbox{\boldmath$\phi$}}
\def\Bchi{\mbox{\boldmath$\chi$}}
\def\Bpsi{\mbox{\boldmath$\psi$}}
\def\Bomega{\mbox{\boldmath$\omega$}}
\def\Bvarepsilon{\mbox{\boldmath$\varepsilon$}}
\def\Bvartheta{\mbox{\boldmath$\vartheta$}}
\def\Bvarpi{\mbox{\boldmath$\varpi$}}
\def\Bvarrho{\mbox{\boldmath$\varrho$}}
\def\Bvarsigma{\mbox{\boldmath$\varsigma$}}
\def\Bvarphi{\mbox{\boldmath$\varphi$}}
\def\bA{\mbox{\boldmath$ A$}}
\def\bB{\mbox{\boldmath$ B$}}
\def\bC{\mbox{\boldmath$ C$}}
\def\bD{\mbox{\boldmath$ D$}}
\def\bE{\mbox{\boldmath$ E$}}
\def\bF{\mbox{\boldmath$ F$}}
\def\bG{\mbox{\boldmath$ G$}}
\def\bH{\mbox{\boldmath$ H$}}
\def\bI{\mbox{\boldmath$ I$}}
\def\bJ{\mbox{\boldmath$ J$}}
\def\bK{\mbox{\boldmath$ K$}}
\def\bL{\mbox{\boldmath$ L$}}
\def\bM{\mbox{\boldmath$ M$}}
\def\bN{\mbox{\boldmath$ N$}}
\def\bO{\mbox{\boldmath$ O$}}
\def\bP{\mbox{\boldmath$ P$}}
\def\bQ{\mbox{\boldmath$ Q$}}
\def\bR{\mbox{\boldmath$ R$}}
\def\bS{\mbox{\boldmath$ S$}}
\def\bT{\mbox{\boldmath$ T$}}
\def\bU{\mbox{\boldmath$ U$}}
\def\bV{\mbox{\boldmath$ V$}}
\def\bW{\mbox{\boldmath$ W$}}
\def\bX{\mbox{\boldmath$ X$}}
\def\bY{\mbox{\boldmath$ Y$}}
\def\bZ{\mbox{\boldmath$ Z$}}
\def\ba{\mbox{\boldmath$ a$}}
\def\bb{\mbox{\boldmath$ b$}}
\def\bc{\mbox{\boldmath$ c$}}
\def\bd{\mbox{\boldmath$ d$}}
\def\be{\mbox{\boldmath$ e$}}
\def\bff{\mbox{\boldmath$ f$}}
\def\bg{\mbox{\boldmath$ g$}}
\def\bh{\mbox{\boldmath$ h$}}
\def\bi{\mbox{\boldmath$ i$}}
\def\bj{\mbox{\boldmath$ j$}}
\def\bk{\mbox{\boldmath$ k$}}
\def\bl{\mbox{\boldmath$ l$}}
\def\bm{\mbox{\boldmath$ m$}}
\def\bn{\mbox{\boldmath$ n$}}
\def\bo{\mbox{\boldmath$ o$}}
\def\bp{\mbox{\boldmath$ p$}}
\def\bq{\mbox{\boldmath$ q$}}
\def\br{\mbox{\boldmath$ r$}}
\def\bs{\mbox{\boldmath$ s$}}
\def\bt{\mbox{\boldmath$ t$}}
\def\bu{\mbox{\boldmath$ u$}}
\def\bv{\mbox{\boldmath$ v$}}
\def\bw{\mbox{\boldmath$ w$}}
\def\bx{\mbox{\boldmath$ x$}}
\def\by{\mbox{\boldmath$ y$}}
\def\bz{\mbox{\boldmath$ z$}}
\title{Elastica-based strain energy functions for soft biological tissue}
\author{K. Garikipati\thanks{Associate Professor, Department of Mechanical Engineering, and
  Program in Applied Physics, University of Michigan, Ann Arbor, USA;
  corresponding author, {\tt krishna@umich.edu}},
  S. G\"oktepe\thanks{Research Associate, Institute of Applied Mechanics, Universit\"{a}t Stuttgart, Germany} \& C. Miehe\thanks{Professor, Dr.-Ing.,
    Institute of Applied Mechanics, Universit\"{a}t Stuttgart,
    Germany}}
\date{University of Michigan \& Universit\"{a}t Stuttgart}
\maketitle

\begin{abstract}
Continuum strain energy density functions are developed for soft biological
tissues that possess slender, fibrillar components. The treatment is based
on the model of an elastica, which is our fine scale model, and is
homogenized in a simple fashion to obtain a continuum strain energy density
function. Notably, we avoid solving the exact, fourth-order, nonlinear,
partial differential equation for deformation of the elastica by
resorting to other 
assumptions, kinematic and energetic, on the response of
individual, elastica-like fibrils. The formulation, discussion of
responses of different models and comparison with experiment are
presented. 
\end{abstract}

\section{Background}
\label{sect1}

Currently-used strain energy density functions for soft biological tissue have
two main origins. Some have been adopted from the
rubber elasticity and polymer elasticity literature, and others have
functional forms that have been chosen to reproduce the characteristic
locking behavior observed in experiments
\citep[see][for a detailed treatment]{Fungbook:93}. Among
the rubber/polymer elasticity 
models are micromechanically-derived ones, which mostly incorporate
entropic elasticity \citep[see][for a
  discussion]{LandLif}. Entropic-elasticity 
models are suitable for materials in which the uncoiling of long chain
molecules under 
axial force causes a decrease in configurational entropy as fewer
configurations become available to the molecule vibrating under its
thermal energy \citep[see][for detailed treatments of these
  models]{Ogdenbook:97}. However, it is not clear that the application
of entropic elasticity is appropriate for many soft biological tissues
such as tendons, ligaments and muscles. As an example, consider the
case of tendons, which have a high collagen
content. \citet{Sunetal:2002} demonstrated by laser trap experiments
that the elasticity of the collagen molecule, which is a triple helix
with a diameter of $1.5$ nm and a contour length (fully uncoiled
length) of approximately $300$ nm, is well-represented by the Worm-like Chain
Model of \cite{KratkyPorod:49}. However, collagen is not restricted
to the form of long chain molecules in the tendon. It
forms fibrils of around $300$ nm diameter, and lengths of the order of
$100$s of $\mu$m. These are further ordered into fibers that can run
the entire length of tendons (the order of cm). The entire
hierarchical structure has extensive crosslinking, including a
longitudinal staggering of the collagen molecules that leads to a
characteristic banded structure on the scale of a fibril, and a
``crimp'' with a wavelength of $10-50$ 
$\mu$m. \citep{Screenetal:2004,ProvenzanoVanderby:2006}. Given this
ordered, hierarchical structure with extensive crosslinking, one must
question the use of entropic elasticity. Due to kinematic constraints
imposed by the crosslinking it seems 
unlikely that the collagen molecules are able to sample many configurations
via thermal fluctuations. Similar arguments can be 
made for ligaments and muscles. 

Support for this view may be inferred from the experiments of
\citet{Wooetal:1987}. Strain-controlled cyclic tension tests of
canine medial collateral ligaments at temperatures between
$2^\circ$ C and $37^\circ$ C showed that the area of hysteresis loops
on the stress-strain diagram
decreased as the temperature of the experiment increased. The use of
strain control implies that the decrease 
in hysteresis was associated with a 
reduction in initial modulus of the stress-strain curve.
In a standard viscoelastic
solid, the initial modulus is a material property; in particular, it
is independent of viscosity, and therefore not subject to mechanisms
of relaxation that may be perceived as a decrease in modulus. This
argument leads to the 
conclusion that, in these experiments, the initial modulus 
of the stress-strain response was decreasing with an increase in
temperature. A decrease in
elasticity with increase in temperature is a signature of
elastic modulus that arises from variation of internal energy, not of
entropic elasticity, which makes the initial modulus increase with 
temperature \citep[see][]{Treloar:1975}.

One reason for the attractiveness of entropic elasticity
models is that they reproduce the experimentally-observed
response of soft biological tissue in uniaxial tension shown in Figure
\ref{fig8}. We draw attention to the characteristics: a prolonged
initial regime with low modulus (the ``toe'' 
region), followed by a short nonlinear regime with rapidly-increasing
modulus (the ``heel'' region) and a final high
modulus region. We will refer to this as the ``characteristic soft
tissue response''. However, there 
are other, non-entropic, models that also reproduce this
behavior. It has been typical in the biomechanics literature to use
strain energy density functions with
mathematical forms that are designed solely to possess this
characteristic response
\citep[see][]{Fungbook:93}. This second class of models is, however, limited
by the lack of microstructural bases for the corresponding strain
energy functions. 

This paper is founded on the recognition that the characteristic locking
behavior can be modelled by an internal energy-based, i.e. non-entropic, model of
elasticity that accounts for the 
uncoiling of crimped fibrils with increasing tension. In these models,
the characteristic soft tissue response is therefore determined by the
elastica-like 
force versus tip displacement response of the individual fibrils. This
consideration 
leads to a 
micromechanically-derived strain energy density function for soft tissue, which,
as argued above, has the proper basis in internal energy, rather
than entropy effects, which are suppressed due to crosslinking.

The realization that characteristic soft tissue response is modelled
by the force-displacement response of an elastica---or approximations of
it---is hardly new. \citet{Diamantetal:72} used a planar model of rigid
links joined by elastic hinges, which they related to the elastica, to
model their observations of stress-stretch behavior of rat tail
tendons. In \citet{Daleetal:72} four kinematic models of crimped fibers were
considered: a planar sinusoidal waveform, a helical shape, a zig-zag
waveform with hinged apices and a zig-zag with apices that undergo
bending to maintain a
constant angle while deforming. The change in profile of these waveforms was
studied and compared with experiment. \citet{BeskosJenkins:75}
modelled  mammalian tendon as an 
incompressible composite with a continuous distribution of
inextensible fibers with a helical shape. The assumption of
inextensibility dominates the response of this model leading to stress
locking in uniaxial tension at a finite stretch. The planar assumption
was also adopted by \citet{ComninouYannas:76}, who modelled single
collagen fibers as sinusoidal beams. Using the theory of shear
deformable beams with linear constitutive relations for
axial stretching and bending, but allowing geometric nonlinearities,
they obtained a nonlinear stress-strain response of
single fibers and extended it to a composite with uniaxial
reinforcement by sinusoidal fibers. In \citet{Lanir:78} a planar model
of a beam on an elastic foundation
was adopted for the mechanical
interaction between collagen (modelled as a beam) and elastin (the
elastic foundation). Similar ideas were explored by
\citet{Kastelicetal:80}, in whose model crimped collagen fibers were
modelled by links that have negligible stiffness until fully
extended. The classical theory of elasticas was used by
\citet{Buckleyetal:80} to treat the deformation of slender
filaments, and the model was solved
numerically. \citet{BasuLardner:85} also studied the stress-stretch 
response of sinusoidal beams in elastic matrices, although they did
not make the link to fibrous soft tissue. A kinematic chain with
finite axial stiffness and torsional springs was used by
\citet{Stoufferetal:85} to represent the uncoiling of crimped fibers,
and compared against experiments. More recently,
\citet{Hurschleretal:97} developed a strain energy density function for tendon
and ligament with seven parameters including microstructural
organization to describe the stress-stretch behavior. The authors also
derived simplified versions of their model that were used to fit
experimentally-determined, nonlinear stress-stretch curves. Finally,
\citet{freed+doehring05} returned to the assumption of a helical
structure for collagen fibrils, and using Castigliano's theorem,
obtained the force-displacement relationship.

In this communication, we present a very general and powerful
procedure for developing the strain energy density function for soft tissue
based on the elastica as a model for slender fibrillar structures. To fix ideas, we
refer to collagen fibrils. In a notable departure from the body of work
cited above, we first obtain the exact, nonlinear, fourth-order
elliptic partial differential equation for the quasi-static deformation
of the extensible elastica. The underlying kinematics are fully
nonlinear and the elastica's strain energy is assumed to be given by
quadratic functions of curvature and the Green-Lagrange strain
tensor.\footnote{The latter dependence is motivated by the
  St. Venant-Kirchhoff model, but there is no further significance to
this choice. Others are equally admissible and do not imply
substantive changes in the outcome.} The difficulty of obtaining
analytic solutions of even simpler 
versions of the
governing partial differential equation has been noted by some of the
authors cited above. Furthermore, numerical solutions, while
possible, will prove both expensive and cumbersome, since our ultimate
aim is a strain energy density function for composite soft tissue in which the
elastica-like fibrils are the reinforcements at a microscopic scale. For
this reason we have examined a few distinct assumptions that make it
possible to obtain force-extension solutions. These
assumptions are related to the kinematic and
energetic behavior of the microscopic collagen fibrils. At the
  outset we wish to emphasize that these assumptions, in addition to being
  well-motivated in our view, are most important to this study because
  they deliver a tractable problem, with only a small number of
  parameters that are also physically-meaningful. This is a theme that we will
  return to at several points in the paper. With this approach we have
  also been able to identify a model that
  can be made to correspond well with experimental data. A rigorous
  validation of these assumptions must, however, await \emph{in
    situ} studies of deforming fibrils. 

The organization of the remainder of the paper is as follows: In
Section \ref{sect2} we lay down  
the fundamental problem of the elastica. The cases of elasticas that
are restricted to circular and sinusoidal arcs, and have further
constraints of kinematics and energetics imposed upon them, are developed in
Sections \ref{eg1} and \ref{sect2b}, respectively. The various
models for the deforming elastica are compared against each other, and
against experiment, in Section \ref{sect2c}. The extension to
macroscopic strain energy density functions, from which the tissue
stress-stretch response can be obtained, and a basic discussion on
convexity appear in Section \ref{sect3}. Closing remarks are made in
Section \ref{closing}.
\section{The deforming elastica}
\label{sect2}

\begin{figure}[ht]
\psfrag{A}{$\be_1$} \psfrag{B}{$\be_2$} \psfrag{C}{$\be_3$}
\psfrag{X}{$\bX$} \psfrag{T}{$\bT$} \psfrag{S}{$S$}
\centering
\includegraphics[width=7cm]{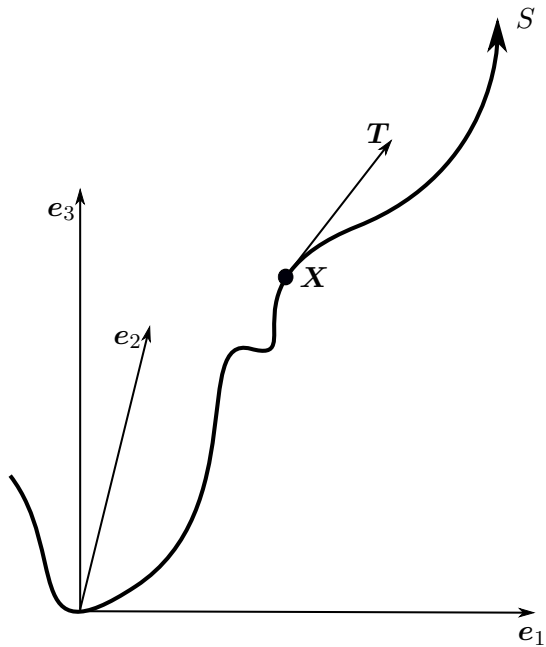}
\caption{The curve, $\Gamma \in \mathbb{R}^3$.}\label{figelastica}
\end{figure}

Consider the elastica, a curve, $\Gamma \subset \mathbb{R}^3$,
parametrized by its arc 
length coordinate, $S$, in the reference configuration (Figure
\ref{figelastica}). Points along the curve are identified by position
vectors $\bX(S) \in \mathbb{R}^3$. The tangent at $S$ is $\bT(S) =
\mathrm{d}\bX/\mathrm{d}S$. Using a Cartesian basis of orthonormal
vectors $\{\be_1,\,\be_2,\,\be_3\}$ it is clear that
$\mathrm{d}X_I/\mathrm{d}S$ is a direction cosine, say $\cos\alpha_I$,
where $I =
1,2,3$, and $\alpha_I$ are the corresponding angles of inclination of
$\bT$. Using the Euclidean norm of a vector, $\Vert\bv\Vert =
\sqrt{\sum_I v_I^2}$, it then follows that  
$\Vert\bT\Vert = 1$. The curvature of $\Gamma$ is $\kappa_0 = \Vert
\mathrm{d}^2\bX/\mathrm{d}S^2\Vert$. In the deformed 
configuration, 
$\gamma$, points have position vectors $\bx(S) = \bX(S) + \bu(S)$. The
tangent vector to $\Gamma$ is carried to $\bt(S) =
\mathrm{d}\bx/\mathrm{d}S$, and by the chain rule it can be written as
\begin{displaymath}
\bt =
\frac{\partial\bx}{\partial\bX}\frac{\mathrm{d}\bX}{\mathrm{d}S}.
\end{displaymath}

\noindent Introducing the deformation gradient tensor, $\bF :=
\partial\bx/\partial\bX$, gives $\bt = \bF\bT$, from which it follows
that $\Vert\bt\Vert^2 = \bT\cdot\bC\bT$, where $\bC =
\bF^\mathrm{T}\bF$ is the right Cauchy-Green tensor. The stretch along
$S$ will be denoted by $\lambda := \mathrm{d}s/\mathrm{d}S =
\Vert\bt\Vert$, and satisfies 
$\lambda > 0$. As an aside, note that if the arc length in the
deformed configuration, say $s$, is used to parameterize $\bx$ as
$\bx(s)$, then the tangent with respect to $\gamma$, which we will
denote by $\bt^{\#}$, is given by $\bt^{\#} = \mathrm{d}\bx/\mathrm{d}s$, and
satisfies $\Vert\bt^{\#}\Vert = 1$. The curvature of $\gamma$ is
$\kappa = \Vert\mathrm{d}^2\bx/\mathrm{d}s^2\Vert$, and, clearly, it can
be written as $\kappa =
\Vert\mathrm{d}^2\bx/\lambda^2\mathrm{d}S^2\Vert$. 

Motivated by the decoupling of the bending and stretching energies of
the classical Euler-Bernoulli beam, we assume that in the present
nonlinear setting the strain energy of $\Gamma$ can also be
decomposed into bending and stretching components, $W(\kappa,\lambda) =
K(\kappa) + U(\lambda)$. The bending stiffness is denoted by
$B$, the stretching modulus by $E$, cross-sectional area by
$A$, and the reference  contour length by $L$. This gives 
\begin{equation}
K(\kappa) = \int\limits_0^L\frac{1}{2}B(\kappa -
\kappa_0)^2\mathrm{d}S,\quad U(\lambda) =
\int\limits_0^L\frac{1}{2}EA(\lambda^2 - 1)^2
\mathrm{d}S.\label{energies}
\end{equation}

\noindent Observe that $W(\kappa,\lambda),\, K(\kappa)$ and
$U(\lambda)$ denote functionals of the corresponding arguments. For
simplicity, we will assume that 
the bending and axial 
stiffnesses are constant; $B = \mathrm{const.}$, $EA =
\mathrm{const.}$ Also see Section \ref{constraints} in this regard.

The governing partial differential equation for the deformation of the
elastica is obtained by imposing stationarity of the following free
energy functional, in which the above definitions of curvature and
stretch have been exploited:

\begin{equation}
\mathscr{G}[\bu] = \int\limits_0^L \left(\frac{1}{2}B\left(\left\Vert
\frac{\mathrm{d}^2\bx}{\mathrm{d}s^2}\right\Vert - \kappa_0(S)\right)^2
 + \frac{1}{2} EA \left(\left\Vert
\frac{\mathrm{d}\bx}{\mathrm{d}S}\right\Vert^2 - 1\right)^2 - \bq\cdot\bu \right)\mathrm{d}S.  \label{gibbsenergy}
\end{equation}

\noindent Note that the strain energy has been incorporated from
(\ref{energies}) and $\bq(S)$ is the external force per unit
contour length. We have also assumed that the displacement is
specified at $S = \{0,L\}$: $\bu(0) = \bzero$ and $\bu(L) =
\bg$. Introducing the variations $\bx_\varepsilon = \bX + \bu + 
\varepsilon\bw$, where $\varepsilon \in \mathbb{R}$ and $\bw(S)$ is an
arbitrary vector field satisfying $\bw(S) = \bzero$ at $S = \{0,L\}$,
the stationarity condition is

\begin{equation}
\frac{\mathrm{d}}{\mathrm{d}\varepsilon}\mathscr{G}[\bu_\varepsilon]\Big\vert_{\varepsilon=0}
  = \frac{\mathrm{d}}{\mathrm{d}\varepsilon}\left(\int\limits_0^L \left(\frac{1}{2}B\left(\left\Vert
\frac{\mathrm{d}^2\bx_\varepsilon}{\mathrm{d}s^2}\right\Vert -
\kappa_0(S)\right)^2 +
 \frac{1}{2} EA \left(\left\Vert
\frac{\mathrm{d}\bx_\varepsilon}{\mathrm{d}S}\right\Vert^2 -
  1\right)^2 - \bq\cdot\bu_\varepsilon\right) \mathrm{d}S \right)_{\varepsilon=0}= 0.
\label{stationarity}
\end{equation}

Standard variational calculus, the
arbitrariness of $\bw$ and $\mathrm{d}\bw/\mathrm{d}S$, as well as the
homogeneity of $\bw$ at $S = \{0,L\}$, yield the following
Euler-Lagrange equations: 

\begin{equation}
B\frac{\mathrm{d}^2}{\lambda\mathrm{d}(\lambda\mathrm{d}S)}\left(\left(1
-
\kappa_0/\kappa\right)\frac{\mathrm{d}^2\bx}{\lambda\mathrm{d}(\lambda\mathrm{d}S)}\right)
- 2EA\frac{\mathrm{d}}{\mathrm{d}S}\left(\left(\lambda^2
-1\right)\frac{\mathrm{d}\bx}{\mathrm{d}S}\right) - \bq = 0, \label{pde}
\end{equation}

\noindent with the boundary conditions
\begin{equation}
\bu(0) = \bzero,\quad \bu(L) = \bg,\qquad
B\left(1 -
\frac{\kappa}{\kappa_0}\right)\frac{\mathrm{d}^2\bx}{\lambda\mathrm{d}(\lambda\mathrm{d}S)} 
= 0
\;\mathrm{at}\; S = \{0,L\}. \label{bcs}
\end{equation}

Since $\lambda =
\Vert\mathrm{d}\bx/\mathrm{d}S\Vert$ and $\kappa =
\Vert\mathrm{d}^2\bx/\lambda^2\mathrm{d}S^2\Vert$, Equation
(\ref{pde}) possesses complexity beyond that apparent in its 
form above. In addition to the boundary conditions on $\bu$ at
$\{0,L\}$, note that the generalized force satisfies homogeneous
boundary conditions at
$S = \{0,L\}$ in (\ref{bcs})$_3$.\footnote{This generalized force is
  conjugate to $\mathrm{d}\bw/\mathrm{d}S$ in the Euler-Lagrange
  equations arising from (\ref{stationarity}), and therefore admits the
interpretation of a moment.} Alternate boundary conditions can be used
in the above variational procedure. 

\subsection{Simplifying assumptions on kinematics and energetic
  response of the elastica}
\label{constraints}

The ultimate aim of this work is the development of a
macroscopic strain energy density function, where the micromechanics arises
from the deformation of the elastica. The \emph{exact} micromechanics
of the deforming elastica is obtained by solving the partial
differential equation (\ref{pde}) subject to the boundary conditions
in (\ref{bcs}). Its solution, however, is nontrivial on account of the
nonlinearity and fourth-order form of the partial differential
equation. It is desirable to avoid the complexity and
expense of solving this equation repeatedly in fine scale computations
that will be coarse-grained in some suitable fashion to obtain the
strain energy density function and stress at each material point on the
macroscopic scale. For this reason, we examine certain assumptions,
kinematic and energetic, which simplify the micromechanics.  

The kinematic assumption in Section \ref{eg1} is of an elastica which
deforms through a family of circular arcs. In Section \ref{sect2b} it
is a family of sinusoidal waveforms. In both sections we also consider
two further kinematic assumptions, inextensibility and planar
incompressibility, and the assumption of a stationary strain
energy. Planar waveforms have been used 
in many of the studies cited in Section \ref{sect1}, and reported in
the experimental studies of 
\cite{Screenetal:2004} and \cite{ProvenzanoVanderby:2006}. Helical
forms have also been reported in a few cases
\citep[see][although even these authors cite at least an equal
  number of papers that reported planar
  waveforms]{BeskosJenkins:75,freed+doehring05}. Since 
there exists some dispute in the literature in this regard, we have
chosen the simplicity of planar waveforms. We also note the ease of
parametrization of the circular and sinusoidal forms. 

The persistence of the initial family of waveforms, and the
assumptions of inextensibility or planar incompressibility are, perhaps, the
strongest assumptions in this paper. The assumption on persistence of
the family of 
waveforms maintains the ease of parametrization gained by assuming
circular arc and sinusoidal initial waveforms. For each of the assumed waveform
families (circular arc and sinusoidal) and additional kinematic
assumptions (inextensibility and planar incompressibility) we treat
the effective elastic parameters, 
$B$ and $E$, as those measured in an experiment in which the elastica's
deformation remains within the waveform family, with the additional
kinematic assumptions holding. We note that the effective elastic parameters so 
measured will be different from those obtained in unconstrained
experiments. For simplicity we assume $B$ and $E$ to be constant
for each combination of waveform family and kinematic
assumption. Also note, however, that these parameters will differ in
each one of these cases. In each case, we will not solve the
unconstrained problem, but will restrict ourselves to the specific
waveform family, and additional kinematic assumptions, with effective elastic
parameters that are assumed to be obtained from corresponding
experiments.   

Aside from the implication of these effective elastic parameters,
we assume that the distributed forces, $\bq$ in (\ref{pde}),
vanish. We will focus on the \emph{external force} that is conjugate to
the displacement boundary conditions in (\ref{bcs}) in the rest of
this paper.

\noindent\textbf{Remark 1}. The treatment of the kinematics discussed
  above is in direct analogy with constrained formulations in
  classical, linearized elasticity, such as the plane strain constraint, and the
  incompressibility constraint. Consider plane strain: The 
  constraint is assumed to be exactly imposed, and we
  only consider strains in a three-dimensional subspace of
  the full, six-dimensional space. Denote the strains and stresses
  by $\varepsilon_{ij}$ and $\sigma_{ij}$, respectively. Consider a
  plane strain problem in which the
  body is loaded by controlling $\varepsilon_{11}$ while letting
  $\sigma_{22}, \sigma_{12} = 0$ and maintaining the plane strain
  constraint, $\varepsilon_{33}, \varepsilon_{13}, \varepsilon_{23} =
  0$. Importantly, the effective elastic modulus obtained for the
  $\sigma_{11}-\varepsilon_{11}$ response
  differs from the unconstrained case in which, also, $\varepsilon_{11}$ is
  controlled while $\sigma_{ij} = 0$ for $i,j \neq 1$. With Young's
  modulus E and Poisson ratio $\nu$ for an isotropic material, this
  effective modulus is $E/(1 - \nu^2)$ 
  in plane strain, while in the unconstrained case it is $E$. In plane
  strain, $\sigma_{33} \neq 0$ is the Lagrange multiplier enforcing
  $\varepsilon_{33} = 0$, and $\sigma_{13}, \sigma_{23} = 0$ are the
  Lagrange multipliers corresponding to $\varepsilon_{13},
  \varepsilon_{23} = 0$, respectively. It will become apparent in
  Sections \ref{eg1}--\ref{sect2b} that the only loading condition of
  interest in this paper is exactly analogous to the uniaxial loading
  discussed in this remark in the plane strain context. The
  identical arguments, point for point, can be made for the
  incompressibility constraint, 
  $\varepsilon_{11} + \varepsilon_{22} + \varepsilon_{33} = 0$.

\section{Force-displacement response of an elastica deforming as a circular arc} 
\label{eg1}

\begin{figure}[ht]
\psfrag{e1}[l][l]{$\be_1$} 
\psfrag{e2}[l][l]{$\be_2$} 
\psfrag{e3}[r][r]{$\be_3$} 
\psfrag{t} [c][c]{$2\theta(g)$} 
\psfrag{T} [c][c]{$2\theta_0$} 
\psfrag{R} [l][l]{$R$} 
\psfrag{r} [r][r]{$r(g)$} 
\psfrag{g} [c][c]{$g$} 
\psfrag{G}[r][r]{$\Gamma$}
\psfrag{H}[l][l]{$\gamma$} 
\centering
\includegraphics[width=9cm]{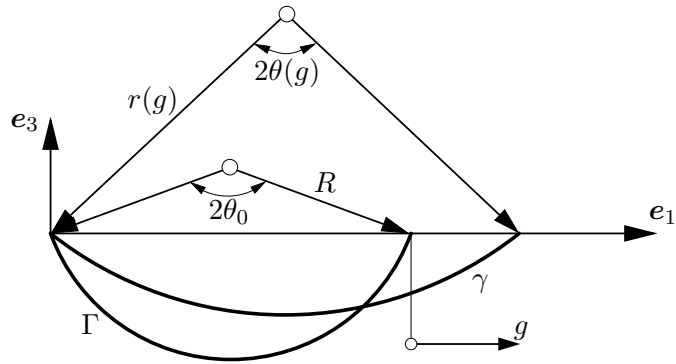}
\caption{Uncoiling of an elastica shaped as a circular arc.}
\label{figeg1} 
\end{figure}

Consider an elastica with reference configuration, $\Gamma$, in the
form of an arc of a circle with central angle $2\theta_0$ and
radius $R$, as shown in Figure \ref{figeg1}. The reference positions
of points on $\Gamma$ are 
\begin{equation}
\bX(S) = \left\{\begin{array}{c} R\left(\sin\theta_0 - \sin(\theta_0 -
S/R)\right)\\ 0 \\
R\left(\cos\theta_0 - \cos(\theta_0 -
S/R)\right)\end{array}\right\},\qquad S \in [0,2R\theta_0]
\label{refx}
\end{equation}

\noindent In Sections 
\ref{circinext}--\ref{eg2} we explore the effect of further
assumptions on this deforming elastica. These are assumptions of
(\romannumeral 1) inextensibility, (\romannumeral 2) planar incompressibility
of the bounding medium, and (\romannumeral 3) stationarity of the
strain energy. These cases are not exhaustive. However, the
physical interpretations and motivations are transparent. 

\subsection{The inextensible elastica deforming as a circular arc}
\label{circinext}

The inextensibility assumption is motivated by the relative stiffness
in the arcwise direction of the collagen fibril in comparison with the
large compliance due to their highly crimped waveform. 

Let the point $S = 2\theta_0 R$ be displaced by the vector $g\be_1$ while the
deformed configuration, $\gamma$,
maintains the form of a circular arc without extension, i.e., $\lambda
= 1$. Then, the tip displacement is restricted to $0 \le g \le 
2(\theta_0 - \sin\theta_0)R$. At a given tip displacement, $g$, the
central angle of the deformed elastica is $\theta(g) =\theta_0
R/r(g)$. The deformed radius, $r(g)$, then 
satisfies the implicit relation
\begin{equation}
r(g)\sin\left(\frac{\theta_0 R}{r(g)}\right) = R\sin\theta_0 + \frac{g}{2}
\label{implicitr}
\end{equation}

\noindent and the positions of points on $\gamma$ are  
\begin{equation}
\bx(S) = \left\{
\begin{array}{c} 
r(g)\left(\sin\theta(g) -\sin\left(\theta(g) - \frac{S}{r(g)}\right)\right)\\ 
0 \\
r\left(\cos\theta(g) -
\cos\left(\theta-\frac{S}{r(g)}\right)\right)\\
\end{array}
\right\}\ .  
\label{defox}
\end{equation}

\noindent From (\ref{refx}) and (\ref{defox}),
\begin{equation}
\frac{\mathrm{d}^2\bX}{\mathrm{d}S^2} =
\left\{
\begin{array}{c}
\frac{1}{R}\cos\left(\frac{S}{R}\right)\\ 
0 \\
\frac{1}{R}\sin\left(\frac{S}{R}\right)
\end{array}
\right\},
\quad
\frac{\mathrm{d}\bx}{\mathrm{d}S} =
\left\{
\begin{array}{c}
\cos\left(\theta(g) - \frac{S}{r(g)}\right)\\ 
0 \\ 
-\sin\left(\theta(g) - \frac{S}{r(g)}\right)\\ 
\end{array}\right\},
\quad
\frac{\mathrm{d}^2\bx}{\mathrm{d}S^2} = 
\left\{
\begin{array}{c}
\frac{1}{r(g)}\sin\left(\theta(g)-\frac{S}{r(g)}\right)\\ 
0 \\ 
\frac{1}{r(g)}\cos\left(\theta(g) - \frac{S}{r(g)}\right)
\end{array}\right\},
\label{elasticakinem} 
\end{equation}

\noindent from which it follows that $\kappa_0 = 1/R$, $\kappa =
1/r$ and $\lambda = 1$. Note that, for $g = 2(\theta_0 -\sin\theta_0)R$,
$\gamma$ is a straight segment of length $2\theta_0 R$ along
$\be_1$. Following (\ref{energies}), the strain energy of the elastica
can now be written as
\begin{equation}
W(\kappa;g) =
   \int\limits_0^{2\theta_0 R}\frac{1}{2}B\left(\kappa(g) -
  \kappa_0\right)^2\mathrm{d}S,\quad 0\le g \le 2(\theta_0 - \sin\theta_0)R
\label{elasticaenergies}
\end{equation}

\noindent Note that in addition to $W(\kappa;g)$ being a functional of the field,
$\kappa$, it is a function of the tip displacement, $g$. The force
response of the elastica to the tip displacement, $g$, is  
\begin{equation}
f(\kappa;g) =
   \frac{\partial W}{\partial g}, \quad 0 \le g \le 2(\theta_0 - \sin\theta_0)R
\label{elasticaforce}
\end{equation}

\noindent Like $W$, the force, $f$, is a functional of $\kappa$ and a
function of $g$. In what follows, the functional character of $f$ is
suppressed since its dependence on $g$ is of primary interest. 
Using $\kappa = \Vert\mathrm{d}^2\bx/\mathrm{d}s^2\Vert$, and equations
(\ref{elasticakinem}), (\ref{implicitr}) and (\ref{elasticaenergies})
we have,   
\begin{equation}
f(g) = \frac{B\theta_0 R}{r(g)^2}
\left(\frac{1}{r(g)} - \frac{1}{R}\right)
\frac{1}{\theta\cos\theta-\sin\theta}
,\quad 0 \le g \le 2(\theta_0 -\sin\theta_0)R.
\label{elasticaforce1}
\end{equation}

Equation
(\ref{elasticaforce1}) indicates that 
$f(g)$ diverges as $1/r(g) \to 0$. Thus the force in a fully uncoiled,
inextensible elastica diverges. An extensible elastica, however, develops finite
axial tension due to stretching along the tangent
as it is uncoiled, and will be considered in the next two sections. 

\noindent\textbf{Remark 2}. We reiterate that the above approach does
not involve a formal  
solution of (\ref{pde}). Instead, it is assumed, \emph{a priori},
that this governing equation is satisfied with the inextensible elastica
maintaining the circular arc form. 

\subsection{The extensible elastica deforming as a circular arc and subject to
  macroscopic, planar incompressibility}   
\label{circincomp}

\begin{figure}[ht]
\psfrag{e1}[l][l]{$\be_1$} 
\psfrag{e2}[l][l]{$\be_2$} 
\psfrag{e3}[r][r]{$\be_3$} 
\psfrag{area0}[l][l]{$2R^2\sin\theta_0(1-\cos\theta_0)$} 
\psfrag{area} [l][l]{$a(2R\sin\theta_0+g)$} 
\psfrag{t} [c][c]{$2\theta(g)$} 
\psfrag{T} [c][c]{$2\theta_0$} 
\psfrag{R} [c][c]{$R$} 
\psfrag{r} [r][r]{$r(g)$} 
\psfrag{a} [c][c]{$a$} 
\psfrag{g}[l][l]{$g$} 
\psfrag{G}[r][r]{$\Gamma$}
\psfrag{H}[l][l]{$\gamma$} 
\centering
\includegraphics[width=9cm]{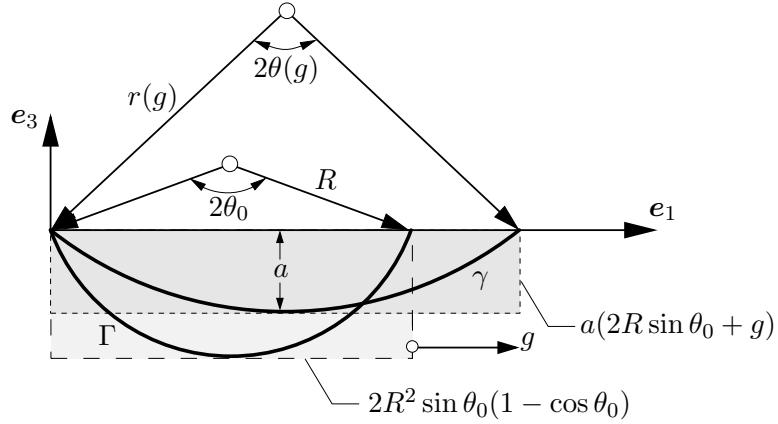}
\caption{A circular arc elastica surrounded by a two-dimensionally
  incompressible medium.} 
\label{cir2} 
\end{figure}

Collagen fibrils in soft tissues are surrounded by proteoglycan
molecules that bind water. At the levels of stress that the tissue is
subject to, the proteoglycan matrix is nearly
incompressible. Motivated thus, we consider the waveform of the deforming fibril
to be subject to planar incompressibility as a model for full
three-dimensional incompressibility (see Remark 5). The elastica
deforming as a circular arc 
in the plane spanned by $\{\be_1,\be_3\}$ is also subject to
invariance of the area of a circumscribing rectangle, even as the
rectangle's aspect ratio varies with tip displacement, $g$. See Figure
\ref{cir2}. The conservation of the area, then, leads us to a
closed--form expression for the height, $a$, of the current rectangle.
\begin{equation}
A_0 \equiv A
\quad\leadsto\quad 
a(g) = \frac{2R^2\sin\theta_0\,(1-\cos\theta_0)}{2R\sin\theta_0+g} \:.
\label{incomp-circ}
\end{equation}

With this explicit form of $a(g)$, the radius of the deforming
elastica can be determined from the geometry of Figure \ref{cir2} as
$r^2 = (r-a)^2 + (R\sin\theta_0+g/2)^2$. This yields
\begin{equation}
r(g) = \frac{1}{2a(g)}
\left(
\left(a(g)\right)^2 + \frac{\left(R^2\sin\theta_0\,(1-\cos\theta_0)\right)^2}{\left(a(g)\right)^2}
\right)\:.
\label{incomp-radius}
\end{equation}
The current curvature $\kappa(g)$ and stretch $\lambda(g)$
of the circular arc elastica then immediately follow as
\begin{equation}
\kappa(g) = \frac{1}{r(g)}
\quad\mathrm{and}\quad 
\lambda(g)=\frac{r(g)\theta(g)}{R\theta_0}
\label{incomp-circ-kine}
\end{equation}
where the implicit relation (\ref{implicitr}) is now replaced by

\begin{equation}
\theta(g)=\sin^{-1}\left(\frac{R\sin\theta_0+g/2}{r(g)}\right).\label{implicitrincomp}
\end{equation}

\noindent See Figure \ref{cir2}. Inextensibility does not hold: 
$\lambda(g) \neq 1$. Using (\ref{incomp-circ-kine}) for $\kappa(g)$
and $\lambda(g)$, and noting that $\theta$ is a function of $r(g)$ and
$g$, we parametrize the strain energy as

\begin{equation}
\overline{W}(r;g) =
   \int\limits_0^{2\theta_0 R}\frac{1}{2}B\left(\frac{1}{r(g)} -
  \frac{1}{R}\right)^2\mathrm{d}S + \int\limits_0^{2\theta_0
   R}\frac{1}{2}EA\left(\left(\frac{r(g)\theta(g)}{R\theta_0}\right)^2 -1\right)^2\mathrm{d}S,
\label{elasticaenergiesincomp}
\end{equation}

\noindent where, as previously, $\overline{W}$ is a functional of $r$
and a function of $g$. The tip force,
$f(g)=\partial\overline{W}/\partial g$, is  

\begin{equation}
\begin{array}{lll}
f(g)&=& 
\theta_0 B R\displaystyle\left(\frac{1}{r} - \frac{1}{R}\right)
\displaystyle\left(-\frac{1}{r^2}\right)
\left(\frac{3R^2\sin\theta_0(1-\cos\theta_0)}{2a^2}- \frac{a^2}{2R^2\sin\theta_0(1-\cos\theta_0)} \right) \\[3Ex]
&+& 
EA \lambda (\lambda^2 - 1)
\displaystyle\left(2\sec\theta + (\theta-\tan\theta)
\left(\frac{3R^2\sin\theta_0(1-\cos\theta_0)}{a^2}-
\frac{a^2}{R^2\sin\theta_0(1-\cos\theta_0)} \right)\right)
\:, \\ 
\end{array} 
\label{incomp-circ-forc}
\end{equation}
where the derivative formulas
$\mathrm{d}r/\mathrm{d}g = 
{3R^2\sin\theta_0(1-\cos\theta_0)}/{4a^2}- {a^2}/{4R^2\sin\theta_0(1-\cos\theta_0)}\;,
\mathrm{d}\theta/\mathrm{d}g = \sec\theta/2r - \tan\theta\mathrm{d}r/r\mathrm{d}g$ and 
$\mathrm{d}\lambda/\mathrm{d}g =
(\frac{1}{2}\sec\theta+(\theta-\tan\theta)\mathrm{d}r/\mathrm{d}g) / (\theta_0 R)$
have been incorporated. 

\noindent\textbf{Remark 3}. Note that the stretch, $\lambda$, defined in
(\ref{incomp-circ-kine})$_2$ is averaged over the reference contour
length; i.e., over the circular arc with central angle
$2\theta_0$. The use of $\lambda$ in the constitutive relation
(\ref{elasticaenergiesincomp}) implies that the axial stiffness,
$EA$, is also homogenized over the reference contour length. 

\noindent\textbf{Remark 4}. For this case of the circular arc-shaped elastica
in a two-dimensionally incompressible 
medium, it is found that $\lambda < 1$ for a certain range of
\emph{macroscopic stretch}, $\bar{\lambda} = 1 +
g/2R\sin\theta_0$ (see Section \ref{sect2c.1}). The incompressibility
constraint causes 
compression of the elastica. If (\ref{pde}) were solved with this
macroscopic incompressibility constraint it would manifest itself as
buckling. We have not attempted to follow the buckled shape of the
elastica. Instead we have solved for 
the value of $\theta_0 = \theta_{0_\mathrm{cr}}$ such that for all
$\theta_0 < \theta_{0_\mathrm{cr}}$ we have
$\mathrm{d}\lambda/\mathrm{d}\bar{\lambda} > 0$, ensuring that
macroscopic stretch translates to microscopic stretch. This root is
$\theta_{0_\mathrm{cr}} \approx 1.342$. Our numerical studies of the circular
arc-shaped elastica in an incompressible medium (Section \ref{sect2c.1})
are restricted to
$\theta_0 > \theta_{0_\mathrm{cr}}$. 

\noindent\textbf{Remark 5}.\footnote{It was pointed out to
  us by a reviewer that the condition in this subsection is properly
  called a ``planar
  incompressibility condition'', which we have now done.} The more
  physically-realistic condition of volumetric incompressibility leads
  to the relation 

\begin{equation}
V_0 \equiv V
\quad\leadsto\quad 
a(g) = \sqrt{\frac{2R^3\sin\theta_0\,(1-\cos\theta_0)^2}{2R\sin\theta_0+g}} \:,
\label{3dincomp-circ}
\end{equation}

\noindent from which the results follow in the same manner as outlined
in this subsection. The results,
  however, are not qualitatively different between the planar and
  volumetric incompressibility conditions.

\subsection{The extensible elastica deforming as a circular arc, and
  relaxing to a stationary strain energy configuration} 
\label{eg2}

Referring to Figure \ref{figeg1}, and persisting with the stretch,
$\lambda(g)$, averaged over the reference contour length as in
(\ref{incomp-circ-kine})$_2$, the central angle, $\theta(g)$, of the
deformed elastica as in (\ref{implicitrincomp}) and $\kappa(g) =
1/r(g)$, the strain energy is still given by (\ref{elasticaenergiesincomp}).

For a given tip displacement, $g$,
corresponding to an applied force, 
$f$, the elastica deforming as a circular arc is now assumed to relax
to a deformed radius, $r(g)$, at which the strain energy is
stationary. The motivation comes from the idea that like long chain
bio-molecules, a collagen fibril also attains an equilibrium state
with respect to its configuration, in addition to deforming under a
tip displacement, $g$. 

\noindent\textbf{Remark 6}. As explained in Section
\ref{constraints}, the force field $\bq 
= \bzero$. Therefore, the corresponding work term does not enter the
stationarity 
calculations. Furthermore, it is assumed that deformation within a
subspace of $\mathbb{R}^3$ for
displacements, $\bu$, manifests
in the effective elastic parameters, $B$ and $E$ for the elastica
deforming as a circular arc. We are interested in stationarity of
strain energy \emph{within this subspace of deformation that the
  elastica is assumed to explore}. That is, the elastica's deformation
is allowed to vary
 only over those configurations allowed while maintaining the circular
arc form for this stationarity calculation. Therefore, the work done by 
Lagrange multipliers that maintain the circular arc shape does not
enter the stationarity calculations. With reference to Remark 1, an
analogous plane strain calculation would seek stationarity of the
strain energy within the strain subspace $\{\varepsilon_{11},
\varepsilon_{22}, \varepsilon_{12}\}$ while maintaining the plane
strain constraint $\varepsilon_{13}, \varepsilon_{23},
\varepsilon_{33} = 0$. Therefore, the Lagrange multipliers,
$\sigma_{13}, \sigma_{23}, \sigma_{33}$, that enforce these
constraints do no work, and do not need to be considered in such
stationarity calculations. Again, the identical arguments, point for
point, can be made
for a calculation seeking stationarity of the strain energy under the
incompressibility constraint; i.e., in the subspace defined by
$\varepsilon_{11} + \varepsilon_{22} + \varepsilon_{33} = 0$.

The radius, $r(g)$, is the solution of the following stationarity
condition:  
\begin{equation}
\frac{\partial \overline{W}}{\partial r} =
2\theta_0 B R\left(\frac{1}{R}- \frac{1}{r}\right)\frac{1}{r^2} 
\ + \ 
4EA \lambda (\lambda^2-1) (\theta - \tan\theta) = 0.
\label{min}
\end{equation}
\noindent
With the deformed radius thus known, the force is 
\begin{equation}
f = \frac{\partial \overline{W}}{\partial g} =
2EA \lambda \left(\lambda^2-1\right)\sec\theta\;,
\label{elasticaforce3}
\end{equation}

\noindent since the contribution to $f$ from the derivatives
$(\partial\overline{W}/\partial r)(\mathrm{d} r/\mathrm{d}g)$
vanishes by (\ref{min}). 

Note that the extension of the above results
(\ref{elasticaforce1}), (\ref{incomp-circ-forc}) 
and (\ref{elasticaforce3}) for half-wavelengths to an elastica
whose waveform consists of $n$ such half-wavelengths is
straightforward. By symmetry, the force, $f$, that results in a total tip
displacement of $g$ corresponds to an extension of $g/n$ of each
half-wavelength, and is obtained from (\ref{elasticaforce1}),
(\ref{incomp-circ-forc}) or
(\ref{elasticaforce3}) by substituting $g$ with $g/n$.


\section{The force-displacement response of a sinusoidal elastica}
\label{sect2b}

\begin{figure}[ht]
\psfrag{e1}[l][l]{$\be_1$} 
\psfrag{e2}[l][l]{$\be_2$} 
\psfrag{e3}[r][r]{$\be_3$} 
\psfrag{L} [c][c]{$l_0$} 
\psfrag{f} [c][c]{$f$} 
\psfrag{g} [c][c]{$g$} 
\psfrag{A} [c][c]{$a_0$} 
\psfrag{a} [c][c]{$a$} 
\centering
\includegraphics[width=8cm]{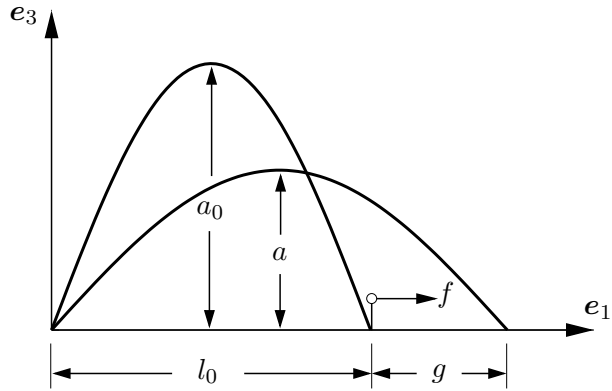}
\caption{Uncoiling of a sinusoidal elastica.}\label{sin1}
\end{figure}

For the sinusoidal waveform, the reference configuration of the
elastica is defined by two shape
parameters, the amplitude $a_0$ and the half--wave length $l_0$.
See Figure \ref{sin1}. 
As in Section \ref{eg1} the elastica lies in the plane
spanned by
$\{\be_1,\be_3\}$. The reference
positions of points on the elastica can be expressed by a single 
parameter, $t$, as 
\begin{equation}
\bX(t) = 
\left\{
\begin{array}{c} 
  X_1(t)\\ 
  0 \\ 
  X_3(t)\\ 
\end{array}
\right\}
=
\left\{
\begin{array}{c} 
  t\\ 
  0 \\ 
  a_0\sin\left(\displaystyle\frac{\pi}{l_0}t\right)\\ 
\end{array}
\right\} \qquad 
\label{X0}
\end{equation}
where $t\:\in [0,l_0]\:.$ Analogously, the shape of the deformed
elastica can also be formulated in terms of a spatial parameter,
$\tilde t$,       
\begin{equation}
\bx(\tilde t) = 
\left\{
\begin{array}{c} 
  x_1(\tilde t)\\ 
  0 \\ 
  x_3(\tilde t)\\ 
\end{array}
\right\}
=
\left\{
\begin{array}{c} 
  \tilde t\\ 
  0 \\ 
  a\sin\left(\displaystyle\frac{\pi}{l}\tilde t\right)\\ 
\end{array}
\right\} \qquad 
\label{x0}
\end{equation}
where $\tilde t\:\in [0,l]\:.$ The deformed
half--wavelength, $l$, is determined by the tip displacement $g$
through the relation $l := l_0 + g$. A linear relation therefore
exists between $\tilde t$ and $t$
\begin{equation}
\tilde t := t\:l/l_0 = t\: (1 + g/l_0)
\quad \textrm{and} \quad 
\frac{\partial\tilde{t}}{\partial t} = l/l_0 = (1 + g/l_0)\;.
\label{1dF}
\end{equation}
It then follows that the arguments of
the Sine functions in (\ref{X0}) and (\ref{x0}) have
the same value, i.e. 
\begin{equation}
\alpha(t) := 
\displaystyle\frac{\pi}{l_0}t = 
\displaystyle\frac{\pi}{l}\tilde t \:.
\label{alpha}
\end{equation}
This geometrical description of the problem allows us
to introduce the two main kinematic variables, namely the curvature,
$\kappa$, and the stretch, $\lambda$, as functions of
derivatives of $X_3$ and $x_3$ with respect to the Lagrangian $t$ and the
Eulerian $\tilde t$ parameters, respectively. For 
the planar reference and current curves parameterized by $t$ and
$\tilde t$, the general curvature formulas given in 
Section \ref{sect2} can be simplified to the forms  
\begin{equation}
\kappa_0(t) := \displaystyle
\frac{X_3^{\prime\prime}}{(1+{X_3^\prime}^{\:2})^{3/2}}\;,
\qquad
\kappa(\tilde t) := \displaystyle
\frac{x_3^{\prime\prime}}{(1+{x_3^\prime}^{\:2})^{3/2}}\:.
\label{kappas}
\end{equation}
The superscript $(\cdot)^\prime$ in (\ref{kappas}) denotes the derivatives
\begin{equation}
\begin{array}{ll} 
X_3^\prime:=\frac{\partial X_3}{\partial t}=
a_0\displaystyle\frac{\pi}{l_0}\cos(\alpha(t))\;, &
X_3^{\prime\prime}:=\frac{\partial^2 X_3}{\partial t^2}=
-a_0\displaystyle\frac{\pi^2}{l_0^2}\sin(\alpha(t))\;,\\[2Ex]
x_3^\prime:=\frac{\partial x_3}{\partial \tilde{t}}= 
a\displaystyle\frac{\pi}{l}\cos(\alpha(t))\;, &
x_3^{\prime\prime}:=\frac{\partial^2 x_3}{\partial \tilde{t}^2}=
-a\displaystyle\frac{\pi^2}{l^2}\sin(\alpha(t))\:.  
\end{array}
\label{derivatives}
\end{equation}
The local stretch is obtained as
$\lambda =\mathrm{d}s/\mathrm{d} S$ where the infinitesimal arc length
measures $\mathrm{d} S$, and $\mathrm{d} s$ are
$\mathrm{d} S:=\sqrt{\mathrm{d}X_1^2 +\mathrm{d}X_3^2}=
\sqrt{1+{X_3^\prime}^{\:2}}\;\mathrm{d}t$  and 
$\mathrm{d} s:=\sqrt{\mathrm{d}x_1^2 +\mathrm{d}x_3^2}=
\sqrt{1+{x_3^\prime}^{\:2}}\;\mathrm{d}\tilde t\;,$ respectively. 
Then, combining these results with
(\ref{1dF})$_2$ yields the stretch expression
\begin{equation}
\lambda:=\mathrm{d} s/\mathrm{d} S = \displaystyle
\frac{\sqrt{1+{x_3^\prime}^{\:2}}} {\sqrt{1+{X_3^\prime}^{\:2}}} \frac{l}{l_0}\:.
\label{sinstretch}
\end{equation}
The basic kinematic variables $\kappa_0,\kappa(t,g,a(g))$ and
$\lambda(t,g,a(g))$ are thus
defined. Note that $\kappa$ and $\lambda$ vary pointwise with $t$ and
are parametrized by $g$ and $a(g)$. The total energy of the
sinusoidal elastica in the 
reference configuration  is
\begin{equation}
\widetilde{W}(\kappa,\lambda;g,a(g))= \int\limits_0^{l_0}\frac{1}{2}B(\kappa(t,g,a(g)) -
\kappa_0(t))^2J(t)\mathrm{d}t + 
\int\limits_0^{l_0}\frac{1}{2}EA(\lambda^2(t,g,a(g)) - 1)^2
J(t)\mathrm{d}t 
\label{sinusoidenergies}
\end{equation}
where $J(t):=\mathrm{d}S/\mathrm{d}t=\sqrt{1+{X_3^\prime}^{\:2}}\:.$ The tip
force $f(g)$, being energy-conjugate to the tip displacement $g$, is
then given by
\begin{equation}
f(g)= \frac{\partial\widetilde{W}(g,a(g))}{\partial g}=
\left.\frac{\partial\widetilde{W}(g,a(g))}{\partial g}\right|_{a} + 
\: 
\frac{\partial\widetilde{W}(g,a(g))}{\partial a}\;
\frac{\mathrm{d}a(g)}{\mathrm{d}g} 
\label{sinusoidforce1}
\end{equation}
or 
\begin{equation}
\begin{array}{lll}
f(g)&=& \displaystyle\int\limits_0^{l_0} 
B(\kappa(t,g,a(g)) - \kappa_0(t))
\left.\frac{\partial\kappa(t,g,a(g))}{\partial g}\right|_{t}
 J(t)\mathrm{d}t
\\
&&+\displaystyle\int\limits_0^{l_0} 
EA(\lambda^2(t,g,a(g)) -
1)
\left.\frac{\partial\lambda^2(t,g,a(g))}{\partial g}\right|_{t}
 J(t)\mathrm{d}t\:. \\ 
\end{array} 
\label{sinusoidforce2}
\end{equation}
\noindent where, as previously, the fact that $f$ and $\widetilde{W}$
are functionals of $\kappa$ and $\lambda$  has been suppressed.

In order to complete the geometric and constitutive description of
the deforming sinusoidal elastica, $a(g)$ must be obtained for
each $g$. Additional kinematic assumptions are made, as for the
elastica deforming as a circular arc. In the case of the sinusoidal
elastica, the additional kinematic assumption determines the current
amplitude $a$. As in Section \ref{eg1} we consider (\romannumeral 1)
inextensibility, (\romannumeral 2) a planar
incompressible bounding medium, and (\romannumeral 3) stationary
strain energy. 

\subsection{Force-displacement response of an inextensible sinusoidal
  elastica} 
\label{sect2b.1}

The local inextensibility condition requires that $\lambda:=\mathrm{d}
s/\mathrm{d} S =1$ and thus  
\begin{equation}
\lambda^2-1= \displaystyle
\frac{{1+{x_3^\prime}^{\:2}}} {{1+{X_3^\prime}^{\:2}}} \frac{l^2}{l^2_0} -1 =0\:.
\label{locinextensibility}
\end{equation}
From (\ref{derivatives}) and (\ref{locinextensibility}) we have,
\begin{equation}
a^2 = a^2_0 - \frac{l^2-l^2_0}{\pi^2\,\cos^2(\alpha)}\:.
\label{locinextensibility2}
\end{equation}
Clearly, the right hand-side of (\ref{locinextensibility2}) can be
negative, and is
unbounded from below in the limit $\alpha \to \pi/2$. Even
for positive values of the right hand-side, i.e. $a^2 > 0$,
$a$ as given by (\ref{locinextensibility2}) varies along the
elastica. This indicates that the requirement that $\lambda = 1$
pointwise along the 
elastica is inconsistent with maintenance of the sinusoidal shape. It
is worth noting that, even physically, it is not clear whether
inextensibility should be applied pointwise to model fibrils that are stiff to
arcwise extension. It may well prove to be more appropriate to
consider inextensibility over a larger length scale. For
this reason, we relax this assumption to a weaker one of
conservation of the length of the sinusoidal elastica for a given tip
displacement $g$. We continue to require that $a = a(g)$ (a function
of $g$ only) and therefore is a parameter for the half wavelength. The
weak inextensibility condition gives
\begin{equation}
\int\limits_s \mathrm{d}s= \int\limits_S \mathrm{d}S 
\quad\leadsto\quad
\int\limits_0^{l_0} (\lambda(t,g,a)-1)\,J(t)\,\mathrm{d}t = 0 \:.
\label{gloinextensibility}
\end{equation}
This condition defines a non-linear residual that can be considered
a function of $a = a(g)$ for given $g$:
\begin{equation}
\sR(a) := 
\int\limits_0^{l_0} (\lambda(t,g,a)-1)\,J(t)\,\mathrm{d}t = 0 
\label{resinext}
\end{equation}
In order to solve (\ref{resinext}) a standard Newton-Raphson
iterative scheme must be employed. Recall that this involves the
linearization of
the residual $\sR(a)$ about $a=\bar a$, 
$\left. \mathrm{Lin}\:\sR(a)\right|_{\bar a} :=  \sR(\bar a) +
(\partial\sR/\partial a)|_{\bar a}(a-\bar a) = 0\;,$ and the solution
of this equation 
for $a$ to get
$ a = \:\bar a - {\sR(\bar a)} / (\partial\sR/\partial a)|_{\bar a}\:.$  
For each given value of tip displacement $g$, this iterative update
scheme is repeated until iterates for $a$ converge to within a
tolerance. Once the value of $a$ is computed, we proceed with the   
computation of the tip force $f$. To this end, we need the sensitivity
of the amplitude $a(g)$ to the tip displacement $g$. It can be
calculated by exploiting the implicit form of the residual, now
written as
$\sR(g,a(g))$ for a general displacement controlled loading
process by writing $\mathrm{d}\sR(g;a(g))/\mathrm{d}g =
(\partial\sR/\partial g)|_a + (\partial\sR/\partial a)
(\mathrm{d} a/\mathrm{d} g) = 0$ yielding $\mathrm{d} a/\mathrm{d} g = -
(\partial\sR/\partial g)/(\partial\sR/\partial a)$. With this
sensitivity in hand, the integrands in (\ref{sinusoidforce2}) can be
computed in a straightforward manner:
\begin{equation}
\begin{array}{rlll}
\displaystyle\frac{\partial\kappa}{\partial g}
=&
\displaystyle\left.\frac{\partial\kappa}{\partial g}\right|_a 
+ 
\left.\frac{\partial\kappa}{\partial a}\right|_g 
\frac{\mathrm{d}a}{\mathrm{d}g} &=&
\displaystyle
\frac{x_3^{\prime\prime}}{(1+{x_3^\prime}^{\:2})^{5/2}}
\left(
\frac{({x_3^\prime}^{\:2}-2)}{l}+
\frac{(1-2\,{x_3^\prime}^{\:2})}{a}
 \frac{\mathrm{d}a}{\mathrm{d}g}
\right )\;,\\[4Ex]
\displaystyle\frac{\partial\lambda^2}{\partial g}
=&
\displaystyle\left.\frac{\partial\lambda^2}{\partial g}\right|_a 
+ 
\left.\frac{\partial\lambda^2}{\partial a}\right|_g 
\frac{\mathrm{d}a}{\mathrm{d}g}
&=&
\displaystyle
\frac{2\,l}{l_0^2\;(1+{X_3^\prime}^{\:2})}
\left(
1 + {x_3^\prime}^{\:2}\frac{l}{a}
\frac {\mathrm{d}a} {\mathrm{d}g}
\right)

\:.
\end{array}
\label{integrandderivatives3}
\end{equation}
   
\subsection{Force-displacement response of a sinusoidal elastica
  subject to macroscopic, planar incompressibility}
\label{sect2b.2}
\begin{figure}[ht]
\psfrag{e1}[l][l]{$\be_1$} 
\psfrag{e2}[l][l]{$\be_2$} 
\psfrag{e3}[r][r]{$\be_3$} 
\psfrag{area0} [r][r]{$a_0 l_0$} 
\psfrag{area} [r][r]{$a  l$} 
\psfrag{L} [c][c]{$l_0$} 
\psfrag{l} [c][c]{$l$} 
\psfrag{A} [c][c]{$a_0$} 
\psfrag{a} [c][c]{$a$} 
\centering
\includegraphics[width=8cm]{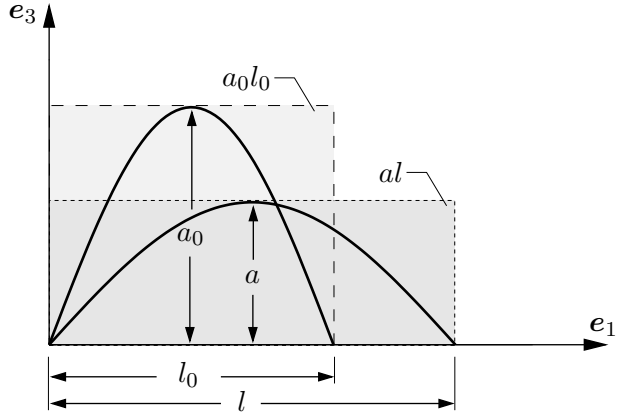}
\caption{A sinusoidal elastica surrounded by an incompressible
  medium.}
\label{sin2} 
\end{figure}

As in Section \ref{circincomp} we use planar incompressibility as a
model for full, three-dimensional incompressibility. The
two-dimensional incompressibility assumption on the surrounding medium
(Figure \ref{sin2}) leads to an explicit result for $a(g)$.  
\begin{equation}
a_0\,l_0 = a\,l \quad \leadsto \quad 
a(g) = \frac{a_0\,l_0}{l} =\frac{a_0\,l_0} {l_0 + g}.
\label{incompressibility}
\end{equation}
Once $a(g)$ is known explicitly in terms of tip displacement $g$, the
derivatives appearing in (\ref{sinusoidforce2}) can be readily
calculated 
\begin{equation}
\frac{\partial\kappa}{\partial g} = 
\frac{3\,x_3^{\prime\prime}}{l}
\frac{({x_3^\prime}^{\:2}-1)}{(1+{x_3^\prime}^{\:2})^{5/2}} 
\quad \mathrm{and}\quad
\frac{\partial\lambda^2}{\partial g}=
\frac{2\,l}{l_0^2}
\frac{(1-{x_3^\prime}^{\:2})}{(1+{X_3^\prime}^{\:2})} \:.
\label{integrandderivatives}
\end{equation}

\subsection{Force-displacement response of a sinusoidal elastica
  with stationary strain energy}
\label{sect2b.3}
For the sinusoidal elastica, stationarity of strain energy is imposed with
respect to the current amplitude $a(g)$, 
i.e. $\partial\widetilde{W}(g,a)/\partial a = 0\:.$ This
condition defines a non--linear residual $\sR(a)$
\begin{equation}
\sR(a) := 
\displaystyle\int\limits_0^{l_0} 
\left( B(\kappa - \kappa_0)
\frac{\partial\kappa}{\partial a} + 
EA(\lambda^2 - 1) \frac{\partial\lambda^2}{\partial a}\right)
 J(t)\mathrm{d}t\:, \\ 
\label{resminen}
\end{equation}
which must vanish for a given tip displacement $g$. As in Section
\ref{sect2b.1} Equation (\ref{resminen}) is solved for $a$ by a
Newton-Raphson iterative scheme. 


With $a$ being known the tip force, $f$, can be computed. Since the
stationary strain energy condition requires that the partial derivative
$\partial\widetilde{W}(g,a)/\partial a$ vanishes, the only terms
contributing to the force will be partial derivatives of kinematic
variables with respect to the tip displacement $g$. That is, it
suffices to compute the integrand terms   
\begin{equation}
\frac{\partial\kappa}{\partial g} = 
\frac{x_3^{\prime\prime}}{l}
\frac{({x_3^\prime}^{\:2}-2)}{(1+{x_3^\prime}^{\:2})^{5/2}} 
\quad \mathrm{and}\quad
\frac{\partial\lambda^2}{\partial g}=
\frac{2\,l}{l_0^2}
\frac{1}{(1+{X_3^\prime}^{\:2})} \:.
\label{integrandderivatives2}
\end{equation}

\noindent\textbf{Remark 7}: Consider the limiting case in which the
tip displacement $g$ is much larger than the
reference half--wavelength $l_0$, i.e. $g/l_0\gg 1$, and 
$l/l_0 \approx g/l_0\gg 1\:.$ 
The elastica tends toward the limiting shape of a straight segment
along $\be_1$. Owing to the flat shape of the elastica, its
spatial slope $x_3^\prime$ and the curvature $x_3^{\prime\prime}$ are
small. This implies that the contribution from the bending 
term to the tip force in both 
(\ref{integrandderivatives})$_1$ and 
(\ref{integrandderivatives2})$_1$
will be negligible in comparison with the contribution from the axial
extension. Furthermore, the vanishing term $x_3^\prime$ in
(\ref{integrandderivatives})$_2$ for large values of the tip
displacement $g$ makes the force terms in 
(\ref{integrandderivatives})$_2$ and 
(\ref{integrandderivatives2})$_2$ tend toward each other. Then, provided the
bending stiffness is not much larger than the axial stiffness, the tip force
values for the planar incompressible bounding medium and stationary energy
cases approach a common value at large tip displacement values. This
is reflected in Figures \ref{fig6} and \ref{fig7}.

\noindent\textbf{Remark 8}: Computations with all the models of the
sinusoidal elastica 
discussed in the preceding Sections \ref{sect2b.1}-- \ref{sect2b.3}
require an efficient numerical integration tool both for
computing the tip force, $f$, and for carrying out the Newton-Raphson
iterations. For this purpose, we have employed a set of F77
subroutines, the so--called {\sc dcuhre}, providing a double
precision integration tool based on adaptive
division of the integration domain into subregions. For details of
the theory and the implementation of the algorithm, the reader is
referred to \citeauthor{berntsen+espelid+genz91a} 
(\citeyear{berntsen+espelid+genz91a}).

\section{Comparison of shape, kinematic and stationary energy assumptions;
  validation}
\label{sect2c}
We now turn to a comparative study of the
force--displacement response of the circular arc and sinusoidal elasticas,
with the additional goal of gaining insight to matches between the
models and experimental data.

\subsection{The force-displacement response of
  circular-arc and sinusoidal elasticas subjected to different
  kinematic assumptions}     
\label{sect2c.1}
\begin{figure}[thb]
\psfrag{0l} [r][r]{\footnotesize  $0$}
\psfrag{1l} [r][r]{\footnotesize  $1$}
\psfrag{2l} [r][r]{\footnotesize  $2$}
\psfrag{3l} [r][r]{\footnotesize  $3$}
\psfrag{4l} [r][r]{\footnotesize  $4$} 
\psfrag{5l} [r][r]{\footnotesize  $5$} 
\psfrag{6l} [r][r]{\footnotesize  $6$} 
\psfrag{7l} [r][r]{\footnotesize  $7$} 
\psfrag{8l} [r][r]{\footnotesize  $8$} 
\psfrag{9l} [r][r]{\footnotesize  $9$} 
\psfrag{10l}[r][r]{\footnotesize $10$}
\psfrag{pi/2l} [r][r]{\tiny $\pi/2$}  
\psfrag{0}    [c][c]{\footnotesize $0$} 
\psfrag{pi/2} [c][c]{\footnotesize $\frac{\pi}{2}$} 
\psfrag{pi}   [c][c]{\footnotesize $\pi$} 
\psfrag{lmax}  [c][c]{\footnotesize 
                      $\bar\lambda_\mathrm{heel}=\theta_{0}/\sin\theta_{0}$} 
\psfrag{theta0}[c][c]{\footnotesize $\theta_0$} 
\psfrag{ 0l}  [r][r]{\footnotesize  $0$} 
\psfrag{ 10l} [r][r]{\footnotesize $10$} 
\psfrag{ 20l} [r][r]{\footnotesize $20$} 
\psfrag{ 30l} [r][r]{\footnotesize $30$} 
\psfrag{ 40l} [r][r]{\footnotesize $40$} 
\psfrag{ 50l} [r][r]{\footnotesize $50$} 
\psfrag{ 1}    [c][c]{\footnotesize $1$} 
\psfrag{ 1.5}  [c][c]{\footnotesize $1.5$} 
\psfrag{ 2}    [c][c]{\footnotesize $2$} 
\psfrag{ 2.5}  [c][c]{\footnotesize $2.5$} 
\psfrag{ pi/2} [c][c]{\footnotesize $\frac{\pi}{2}$} 
\psfrag{f}    [c][c]{\footnotesize Tip force $f$} 
\psfrag{lmac} [c][c]{\footnotesize $\bar\lambda$} 
\psfrag{B1}    [l][l]{\scriptsize $B=1$} 
\psfrag{R1}    [l][l]{\scriptsize $R=1$} 
\psfrag{the030} [l][l]{\footnotesize
  $\theta_0{=}\displaystyle\frac{\pi}{6}$} 
\psfrag{the060} [l][l]{\footnotesize
  $\theta_0{=}\displaystyle\frac{\pi}{3}$}  
\psfrag{the090} [l][l]{\footnotesize
  $\theta_0{=}\displaystyle\frac{\pi}{2}$}  
\psfrag{the0120}[r][r]{\footnotesize
  $\theta_0{=}\displaystyle\frac{2\pi}{3}$}  
\psfrag{a} [l][l]{\normalsize $a$)} 
\psfrag{b} [l][l]{\normalsize $b$)} 
\centering
\includegraphics[width=14cm]{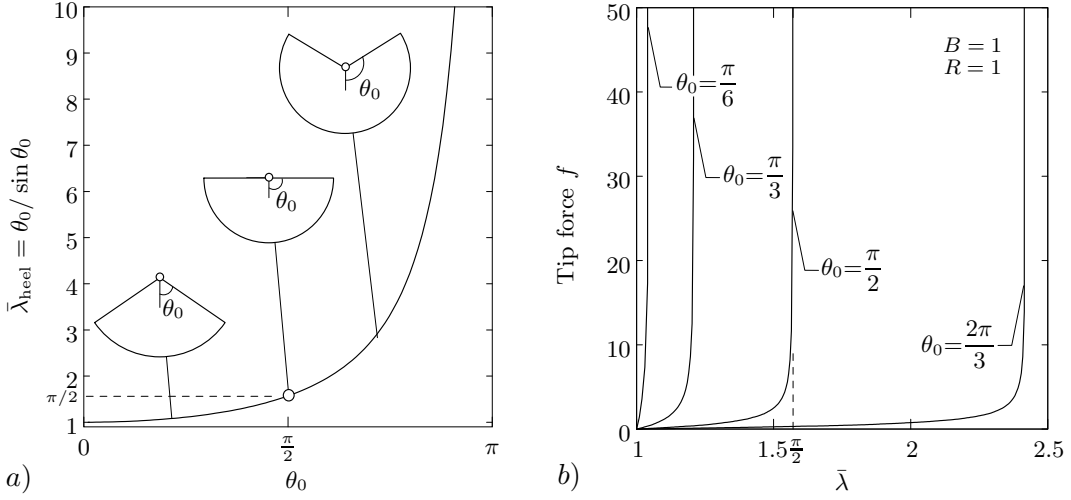}
\caption{Circular-arc, inextensible elastica. $a$) Variation of the maximum 
  macro-stretch $\bar\lambda_\mathrm{heel}=\theta_{0}/\sin\theta_{0}$ with
  the initial angle $\theta_{0}$. $b$) Shift of the maximum value of
  the attainable 
  macro-stretch $\bar\lambda_\mathrm{heel}$ for selected values of the
  initial angle
  $\theta_{0}=\frac{\pi}{6},\frac{\pi}{3},\frac{\pi}{2},\frac{2\pi}{3}$,
  and for $B = 1$, $R = 1$.
}
\label{fig0}
\end{figure}

We first consider the force--stretch behavior of the circular-arc elastica
subject to the two additional kinematic assumptions and the stationary
strain energy assumption. To this end, we first relate the micro--tip
displacement, $g$, and the 
macro--stretch, $\bar\lambda$. The displacement between the ends of the
elastica is assumed to be dictated by macroscopic deformation in an
affine manner. That is, the macro--stretch $\bar\lambda$ is related to
the tip displacement $g$ by $\bar\lambda:= 1+  g/(2
R\sin\theta_{0})$ (see Figure \ref{figeg1}).
\footnote{For tissues with transverse isotropy, where the collagen fibrils
  (elasticas) are characterized by end-to-end vectors that are highly
  aligned, affinity of deformation is a good assumption. The
  alternative, fibril slippage, will be treated in a separate paper.}
In the studies to follow, the macro stretch will be used as the
  primary deformation variable controlling the force.

\begin{figure}[ht]
\psfrag{ 0l}  [r][r]{\footnotesize  $0$} 
\psfrag{ 20l} [r][r]{\footnotesize $20$} 
\psfrag{ 40l} [r][r]{\footnotesize $40$} 
\psfrag{ 60l} [r][r]{\footnotesize $60$} 
\psfrag{ 80l} [r][r]{\footnotesize $80$} 
\psfrag{ 100l}[r][r]{\footnotesize $100$} 
\psfrag{ 1}   [c][c]{\footnotesize $1$} 
\psfrag{ 1.1} [c][c]{\footnotesize $1.1$} 
\psfrag{ 1.2} [c][c]{\footnotesize $1.2$} 
\psfrag{ 1.3} [c][c]{\footnotesize $1.3$} 
\psfrag{ 1.4} [c][c]{\footnotesize $1.4$} 
\psfrag{ 1.5} [c][c]{\footnotesize $1.5$} 
\psfrag{ 1.6} [c][c]{\footnotesize $1.6$} 
\psfrag{f} [c][c]{\footnotesize Tip force $f$} 
\psfrag{l} [l][l]{\footnotesize $\bar\lambda$} 
\psfrag{R0.4} [l][l]{\scriptsize $R=0.4$} 
\psfrag{R0.6} [l][l]{\scriptsize $R=0.6$} 
\psfrag{R0.8} [l][l]{\scriptsize $R=0.8$} 
\psfrag{R1}   [l][l]{\scriptsize $R=1$} 
\psfrag{B1}   [l][l]{\scriptsize $B=1$} 
\psfrag{B4}   [l][l]{\scriptsize $B=4$} 
\psfrag{B7}   [l][l]{\scriptsize $B=7$} 
\psfrag{B10}  [l][l]{\scriptsize $B=10$} 
\psfrag{a} [l][l]{\normalsize $a$)} 
\psfrag{b} [l][l]{\normalsize $b$)} 
\centering
\includegraphics[width=14cm]{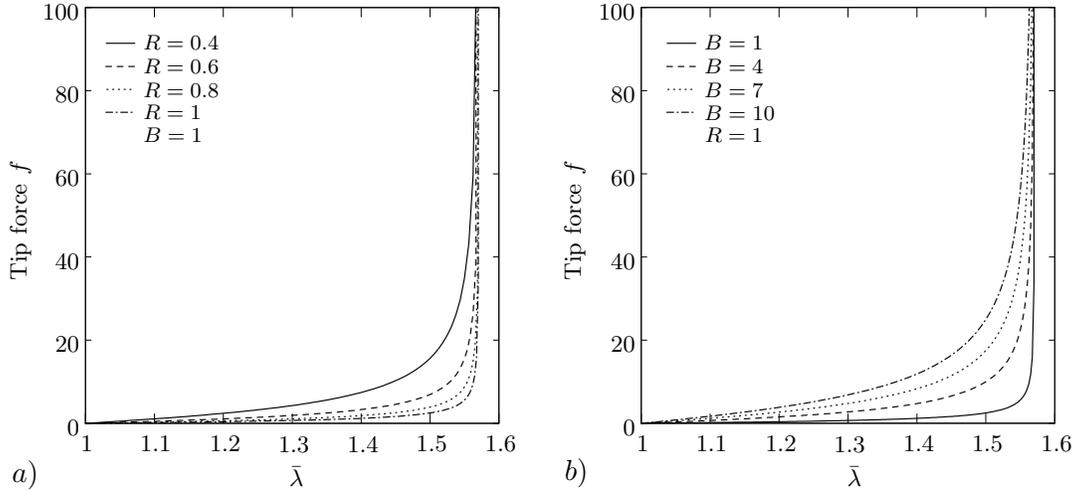}
\caption{Circular-arc. inextensible elastica. Sensitivity analysis of the
  $f-\bar\lambda$ curve to $a$) the variation of initial radius
  $R\in[0.4,1]$ and $b$) the variation of bending stiffness
  $B\in[1,10]$ for $\theta_{0}=\pi/2$. 
}
\label{fig1}
\end{figure}

For the circular-arc, inextensible elastica, according to
(\ref{elasticaforce1}), the tip force $f$ diverges
as the tip displacement $g$ approaches the value
$g_\mathrm{max}=2R(\theta_{0}-\sin\theta_{0})$. This implies that the 
maximum value of macro-stretch is
$\bar\lambda_\mathrm{heel}=\theta_{0}/\sin\theta_{0}\;,$ as shown also in
Figures \ref{fig0}a and \ref{fig0}b. The geometric parameter
$\theta_{0}$ thus has an unambiguous physical effect. The ordinates of
Figure \ref{fig0}a can be obtained as the locking stretch for each value of
$\theta_{0}$ as shown more transparently in Figure \ref{fig0}b for the 
$f{-}\bar\lambda$ response parametrized by $\theta_0$. For the circular-arc,
inextensible elastica, the shape of the $f{-}\bar\lambda$ curve
depends only on the initial radius $R$ and the bending stiffness
$B$. Figures \ref{fig1}a and \ref{fig1}b demonstrate that the
larger the initial radius $R$ or the smaller the bending stiffness
$B$, the sharper is the transition to divergence in $f$. Although
the sharpness of the transition can be tuned, the value of the
locking stretch remains
$\bar\lambda_\mathrm{heel}=\theta_0/\sin\theta_{0}$, and the
response beyond the heel region is asymptotic to a vertical line at
$\bar{\lambda}_\mathrm{heel}$. Of course, this divergent
$f{-}\bar\lambda$ response is non-physical, and considerably limits
the ability to match experiments on different collagenous materials possessing
distinct, and non-divergent, responses in the post-heel region.

\begin{figure}[ht]
\psfrag{ 0l}  [r][r]{\footnotesize  $0$} 
\psfrag{ 20l} [r][r]{\footnotesize $20$} 
\psfrag{ 40l} [r][r]{\footnotesize $40$} 
\psfrag{ 60l} [r][r]{\footnotesize $60$} 
\psfrag{ 80l} [r][r]{\footnotesize $80$} 
\psfrag{ 100l}[r][r]{\footnotesize $100$} 
\psfrag{ 1l}  [r][r]{\footnotesize  $1$} 
\psfrag{ 1.1l}  [r][r]{\footnotesize  $1.1$} 
\psfrag{ 1.2l}  [r][r]{\footnotesize  $1.2$} 
\psfrag{ 1}   [c][c]{\footnotesize $1$} 
\psfrag{ 1.1} [c][c]{\footnotesize $1.1$} 
\psfrag{ 1.2} [c][c]{\footnotesize $1.2$} 
\psfrag{ 1.3} [c][c]{\footnotesize $1.3$} 
\psfrag{ 1.4} [c][c]{\footnotesize $1.4$} 
\psfrag{ 1.5} [c][c]{\footnotesize $1.5$} 
\psfrag{ 1.6} [c][c]{\footnotesize $1.6$} 
\psfrag{ 1.8} [c][c]{\footnotesize $1.8$} 
\psfrag{ 2} [c][c]{\footnotesize $2$} 
\psfrag{ 3} [c][c]{\footnotesize $3$} 
\psfrag{ 4} [c][c]{\footnotesize $4$} 
\psfrag{ 5} [c][c]{\footnotesize $5$} 
\psfrag{f} [c][c]{\footnotesize Tip force $f$} 
\psfrag{l} [l][l]{\footnotesize $\bar\lambda$} 
\psfrag{lam} [c][c]{\footnotesize $\lambda=r\theta/(R\theta_{0})$ } 
\psfrag{EA1}   [l][l]{\scriptsize $EA=1$} 
\psfrag{EA34}  [l][l]{\scriptsize $EA=34$} 
\psfrag{EA67}  [l][l]{\scriptsize $EA=67$} 
\psfrag{EA100} [l][l]{\scriptsize $EA=100$} 
\psfrag{B1}   [l][l]{\scriptsize $B=1$} 
\psfrag{R1}   [l][l]{\scriptsize $R=1$} 
\psfrag{the0} [l][l]{\scriptsize $\theta_{0}=\pi/3$} 
\psfrag{the030} [r][r]{\footnotesize $\theta_0{=}\frac{\pi}{6}$} 
\psfrag{the060} [l][l]{\footnotesize $\theta_0{=}\frac{\pi}{3}$}  
\psfrag{the077} [l][l]{\footnotesize $\theta_{0_\mathrm{cr}}{\approx}1.342$}  
\psfrag{app}    [l][l]{\tiny }  
\psfrag{the090} [l][l]{\footnotesize $\theta_0{=}\frac{\pi}{2}$}  

\psfrag{a} [l][l]{\normalsize $a$)} 
\psfrag{b} [l][l]{\normalsize $b$)} 
\centering
\includegraphics[width=14cm]{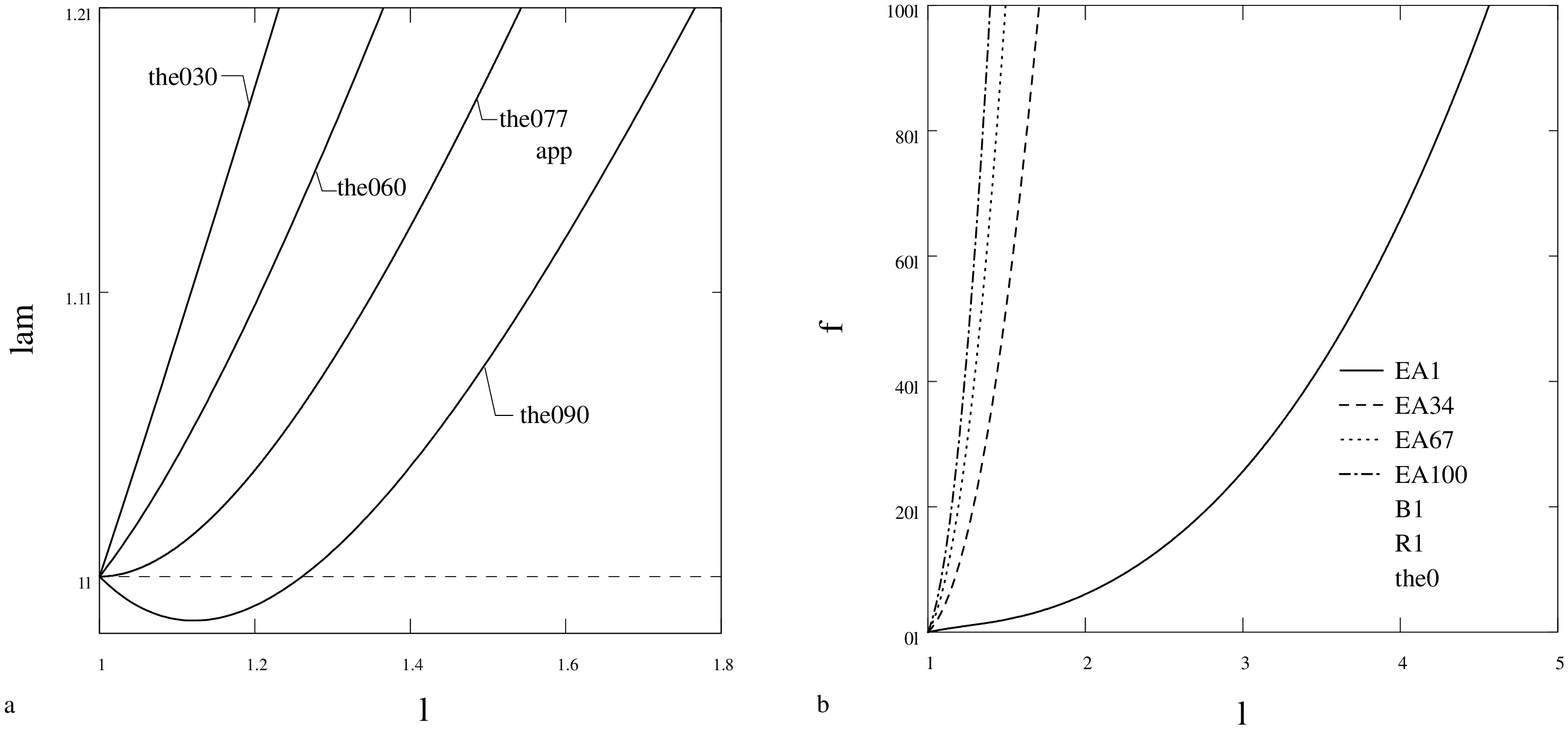}
\caption{Circular-arc elastica surrounded by a planar incompressible
  medium. $a$) Dependence of the micro--stretch $\lambda$
  on the macro--stretch $\bar\lambda$ and the initial angle $\theta_{0}$. 
  $b$) Sensitivity analysis of the $f{-}\bar\lambda$ curve to
  the variation of the axial stiffness  $EA\in[1,100]\;.$
  curve.
}
\label{fig2}
\end{figure}

The variation of the micro-stretch, $\lambda$, with macro-stretch,
$\bar\lambda$, for a 
circular-arc elastica embedded in a planar incompressible medium  is
depicted in Figure \ref{fig2}a. We 
draw attention to the compression of the
elastica for a regime of 
deformation characterized by small values of $\bar{\lambda}$, and
discussed in Remark 4. Following the approach outlined there we have
solved for $\theta_0 = \theta_{0_\mathrm{cr}}$ such that for all $\theta_0 <
\theta_{0_\mathrm{cr}}$ we have
$\mathrm{d}\lambda/\mathrm{d}\bar{\lambda} > 0$. 
Explicitly we have
$\lambda=\theta(\bar\lambda)r(\bar\lambda)/(R\theta_0)$ with  
$r(\bar\lambda)=R(1-\cos(\theta_{0}))/(2\bar\lambda) +
\bar\lambda^{3}R\sin^{2}(\theta_{0})/(2(1-\cos(\theta_{0}))$ and 
$\theta(\bar\lambda)=\sin^{-1}(\bar\lambda
R\sin\theta_{0}/r(\bar\lambda))$ . Setting the derivative
$\mathrm{d}\lambda/\mathrm{d}\bar\lambda|_{\bar\lambda=1}=0$ and
solving the resulting equation for $\theta_{0}$, we obtained the maximum: $\theta_{0}\le  \theta_{0_\mathrm{cr}} \approx 1.342$. As
clearly shown in Figure 
\ref{fig2}a, for the values of the initial angle
$\theta_{0}\le \theta_{0_\mathrm{cr}}$ the compressive regime of
micro-stretch is avoided. For this reason our subsequent
investigations are restricted to 
$\theta_{0}\le \theta_{0_\mathrm{cr}}$ for the circular-arc elastica
embedded in a planar incompressible medium.    

The sensitivity of the tip force, $f$, to the axial stiffness,
$EA$, for $B = 1$,  $R = 1$ and $\theta_{0}=\pi/3$ is depicted in
Figure \ref{fig2}b for the circular-arc elastica surrounded by a planar
incompressible medium. A decrease in the axial 
stiffness translates, as expected, to a decrease of the slope. Clearly, the axial
stiffness dominates the slope of the $f{-}\bar{\lambda}$ curve. 
In contrast to the inextensible case we observe neither a long
toe region nor a distinguishable heel region. Larger values of
$\bar\lambda$ are attainable. As the axial 
stiffness is increased, however, the  $f{-}\bar{\lambda}$ curve enters
the high (but still increasing) slope regime about $\bar{\lambda}_\mathrm{heel}
=\theta_0/\sin\theta_{0}$ without exhibiting much of a toe region. This, of
course, limits the use of this model.    

\begin{figure}[ht]
\psfrag{ 0l}  [r][r]{\footnotesize  $0$} 
\psfrag{ 20l} [r][r]{\footnotesize $20$} 
\psfrag{ 40l} [r][r]{\footnotesize $40$} 
\psfrag{ 60l} [r][r]{\footnotesize $60$} 
\psfrag{ 80l} [r][r]{\footnotesize $80$} 
\psfrag{ 100l}[r][r]{\footnotesize $100$} 
\psfrag{ 1l}  [r][r]{\footnotesize  $1$} 
\psfrag{ 1.1l}  [r][r]{\footnotesize  $1.1$} 
\psfrag{ 1.2l}  [r][r]{\footnotesize  $1.2$} 
\psfrag{ 1}   [c][c]{\footnotesize $1$} 
\psfrag{ 1.1} [c][c]{\footnotesize $1.1$} 
\psfrag{ 1.2} [c][c]{\footnotesize $1.2$} 
\psfrag{ 1.3} [c][c]{\footnotesize $1.3$} 
\psfrag{ 1.4} [c][c]{\footnotesize $1.4$} 
\psfrag{ 1.5} [c][c]{\footnotesize $1.5$} 
\psfrag{ 1.6} [c][c]{\footnotesize $1.6$} 
\psfrag{ 1.8} [c][c]{\footnotesize $1.8$} 
\psfrag{ 2} [c][c]{\footnotesize $2$} 
\psfrag{ 3} [c][c]{\footnotesize $3$} 
\psfrag{ 4} [c][c]{\footnotesize $4$} 
\psfrag{ 5} [c][c]{\footnotesize $5$} 
\psfrag{f} [c][c]{\footnotesize Tip force $f$} 
\psfrag{l} [l][l]{\footnotesize $\bar\lambda$} 
\psfrag{lam} [c][c]{\footnotesize $\lambda=2r\theta/(\pi R)$ } 
\psfrag{EA1}   [l][l]{\scriptsize $EA=1$} 
\psfrag{EA34}  [l][l]{\scriptsize $EA=34$} 
\psfrag{EA67}  [l][l]{\scriptsize $EA=67$} 
\psfrag{EA100} [l][l]{\scriptsize $EA=100$} 
\psfrag{B1}   [l][l]{\scriptsize $B=1$} 
\psfrag{R1}   [l][l]{\scriptsize $R=1$} 
\psfrag{the0} [l][l]{\scriptsize $\theta_{0}=\frac{\pi}{2}$} 
\psfrag{a} [l][l]{\normalsize $a$)} 
\psfrag{b} [l][l]{\normalsize $b$)} 
\centering
\includegraphics[width=7cm]{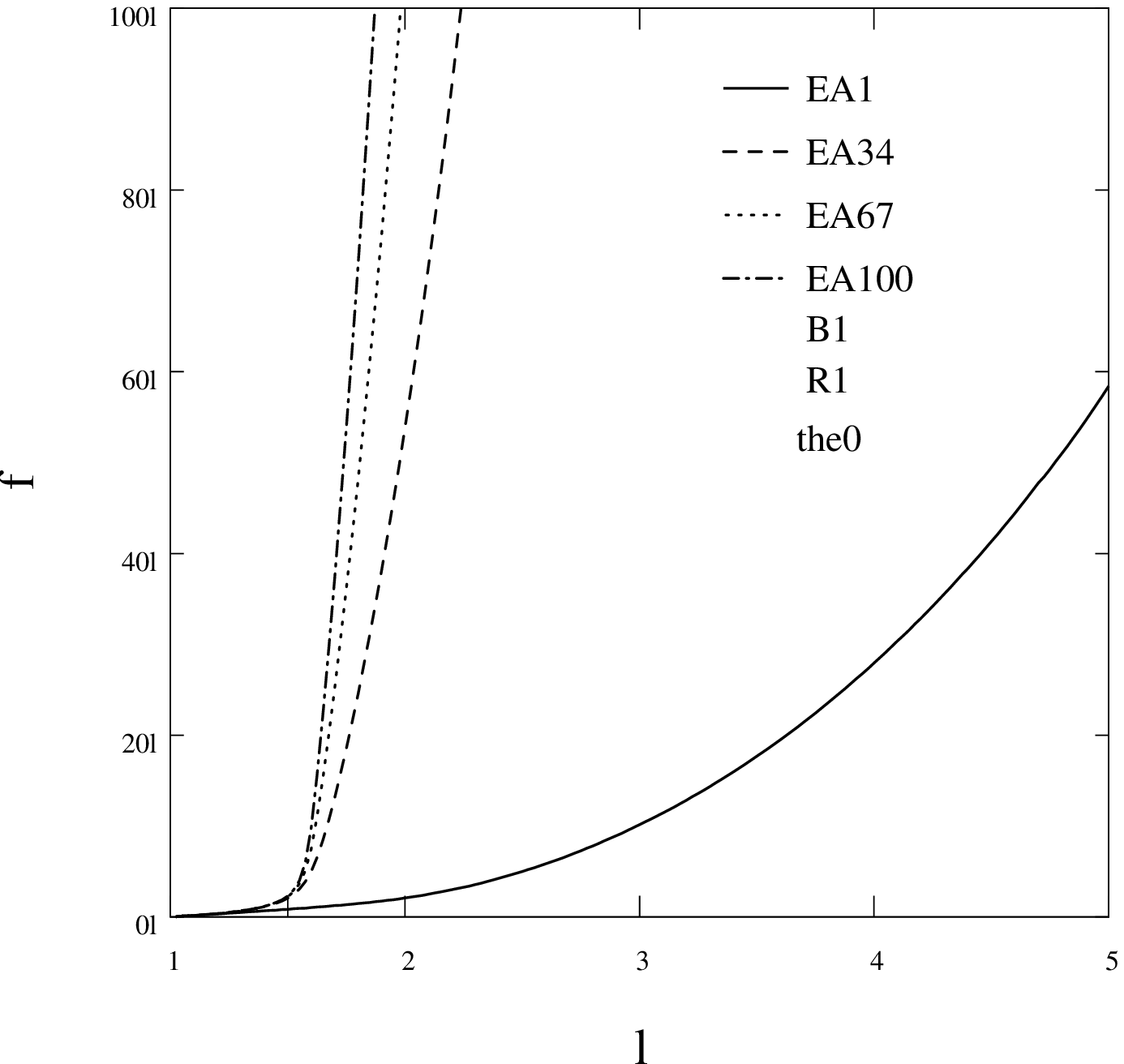}
\caption{Circular-arc elastica with stationary energy. Sensitivity analysis
  of the $f{-}\bar\lambda$ curve to the variation of the axial stiffness
  $EA\in[1,100]\;.$   
}
\label{fig3}
\end{figure}

The material parameters used for the circular-arc elastica
with stationarity of strain energy were the same as for the planar
incompressible medium case. As in the planar incompressible case a decrease in
axial stiffness has a strong, depressing influence on slope of the
$f{-}\bar{\lambda}$ response. The elastica can be extended to
$\bar{\lambda}>\theta_{0}/\sin\theta_{0}\;.$ 
With an increase in axial stiffness, the $f{-}\bar\lambda$ curve approaches
the behavior of the inextensible circular-arc elastica (see Figure
\ref{fig3}). The circular-arc elastica attaining a
stationary strain energy possesses a number of favorable properties:
The toe region exists and its slope can be tuned
by the bending stiffness. The location of the heel region is uniquely
determined by the initial angle $\theta_{0}$ as $\bar{\lambda}_\mathrm{heel}
=\theta_0/\sin\theta_{0}$. The slope of the
post-heel region can be adjusted by the axial stiffness $EA$ as shown in
Figure \ref{fig3}. The stationary strain energy assumption with clearly
identifiable parameters thus serves as a promising model to match with
experimental data.

\begin{figure}[ht]
\psfrag{ 0l}  [r][r]{\footnotesize  $0$} 
\psfrag{ 20l} [r][r]{\footnotesize $20$} 
\psfrag{ 40l} [r][r]{\footnotesize $40$} 
\psfrag{ 60l} [r][r]{\footnotesize $60$} 
\psfrag{ 80l} [r][r]{\footnotesize $80$} 
\psfrag{ 100l}[r][r]{\footnotesize $100$} 
\psfrag{ 1l}  [r][r]{\footnotesize  $1$} 
\psfrag{ 1.1l}  [r][r]{\footnotesize  $1.1$} 
\psfrag{ 1.2l}  [r][r]{\footnotesize  $1.2$} 
\psfrag{ 1}   [c][c]{\footnotesize $1$} 
\psfrag{ 1.1} [c][c]{\footnotesize $1.1$} 
\psfrag{ 1.2} [c][c]{\footnotesize $1.2$} 
\psfrag{ 1.3} [c][c]{\footnotesize $1.3$} 
\psfrag{ 1.4} [c][c]{\footnotesize $1.4$} 
\psfrag{ 1.5} [c][c]{\footnotesize $1.5$} 
\psfrag{ 1.6} [c][c]{\footnotesize $1.6$} 
\psfrag{ 1.8} [c][c]{\footnotesize $1.8$} 
\psfrag{ 2} [c][c]{\footnotesize $2$} 
\psfrag{ 3} [c][c]{\footnotesize $3$} 
\psfrag{ 4} [c][c]{\footnotesize $4$} 
\psfrag{ 5} [c][c]{\footnotesize $5$} 
\psfrag{f} [c][c]{\footnotesize Tip force $f$} 
\psfrag{l} [l][l]{\footnotesize $\bar\lambda$} 
\psfrag{lam} [c][c]{\footnotesize $\lambda=2r\theta/(\pi R)$ } 
\psfrag{EA}   [l][l]{\scriptsize $EA{=}34,67,100$} 
\psfrag{ie} [r][r]{\tiny Inext.} 
\psfrag{me} [r][r]{\tiny Stat. Energy } 
\psfrag{ic} [r][r]{\tiny Incomp.} 
\psfrag{B1}   [l][l]{\scriptsize $B{=}1$} 
\psfrag{R1}   [l][l]{\scriptsize $R{=}1,\:\theta_{0}{=}\pi/3$} 
\centering
\includegraphics[width=7cm]{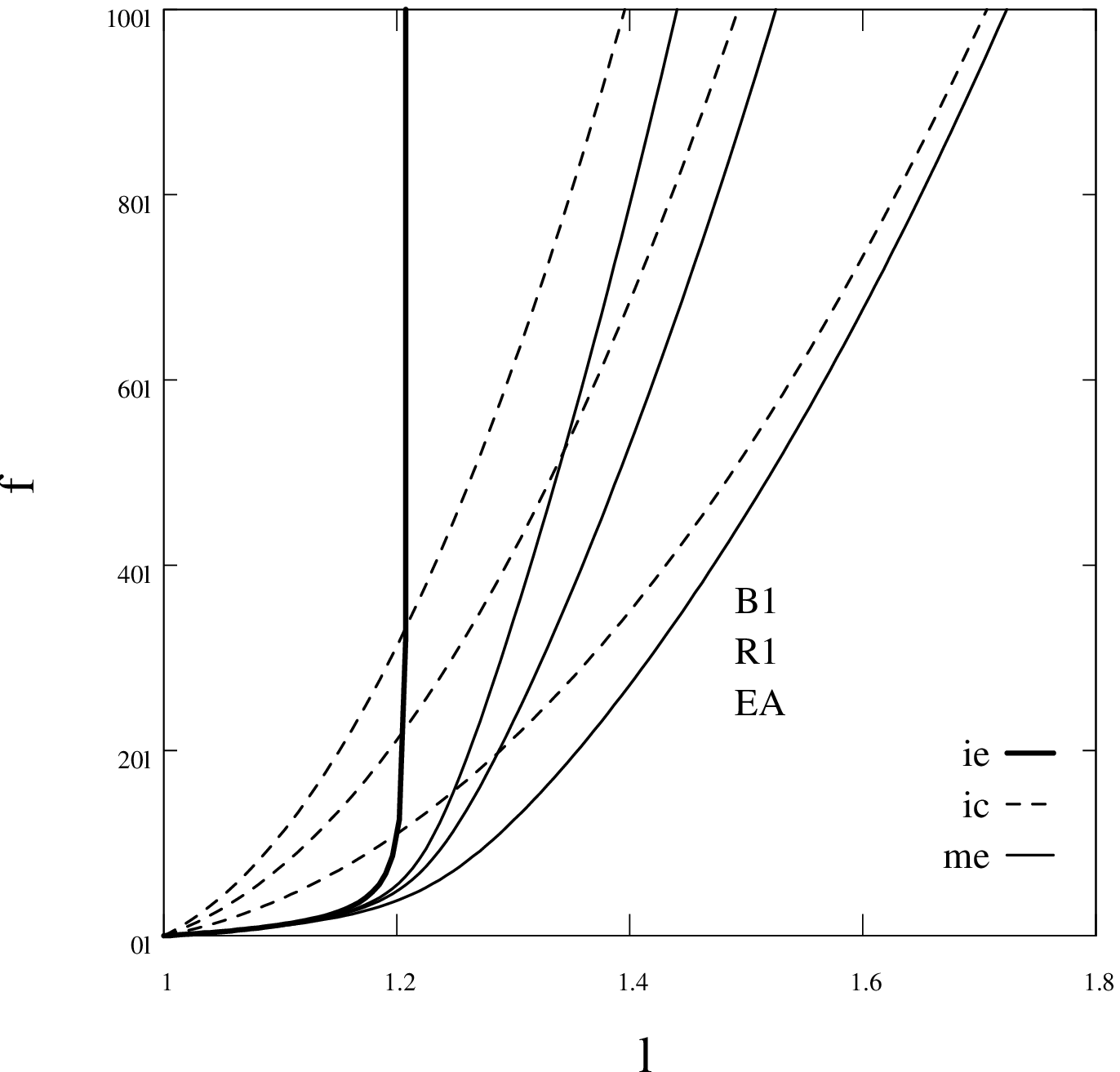}
\caption{Comparison of circular-arc elasticas subjected to different
  constraints. In the planar incompressible and stationary strain energy cases,
  three different values are assigned to the axial stiffness
  $EA=34,67,100$, corresponding to increasingly stiff
  $f{-}\bar{\lambda}$ response.
}
\label{fig4}
\end{figure}

In Figure \ref{fig4} we compare all three cases of the circular-arc
elastica. Distinct values $EA=34,67,100$ are assigned to the axial
modulus of the planar incompressible and stationary energy elasticas. The
$f{-}\bar\lambda$ curve of the stationary energy case approaches the
inextensible one by ``rotating'' about the heel just below
$\bar{\lambda}_\mathrm{heel}$. However, this occurs with no discernible 
difference in the curves for $\lambda$ values smaller than the
heel. In case of the elastica surrounded by a planar incompressible medium,
however, the stiffening in the $f{-}\bar\lambda$ behavior is
different. Owing to larger values of micro--stretch in the initial
stages, the location of the heel shifts to smaller values of
$\bar{\lambda}$.

\noindent
In the foregoing parameter study, we solely considered circular-arc
elasticas with two kinematic assumptions and the stationary strain
energy assumption. In what
follows, we present an analogous parameter sensitivity study for 
the sinusoidal geometry. In contrast to the circular-arc elastica, the
reference shape of a sinusoidal elastica is governed by two
parameters: the amplitude $a_0$ and the half--wave 
length $l_0$ (see Figure \ref{sin1}). The ratio $a_0/l_0$, however, cannot be
arbitrarily chosen. According to the results reported by
\citet{Daleetal:72}, this ratio is limited to 
values smaller than $0.1$. Accounting for this fact in 
the studies to follow the ratio has been chosen as
$a_0/l_0< 0.2\;,$ which will allow us to consider values slightly
larger than the experimental observations. The macro--stretch, $\bar\lambda$,
remains the primary deformation measure, and is now related
to the tip displacement, $g$, by $\bar\lambda = 1+  g/l_0$.

\begin{figure}[ht]
\psfrag{ 0l}  [r][r]{\footnotesize  $0$} 
\psfrag{ 20l} [r][r]{\footnotesize $20$} 
\psfrag{ 40l} [r][r]{\footnotesize $40$} 
\psfrag{ 60l} [r][r]{\footnotesize $60$} 
\psfrag{ 80l} [r][r]{\footnotesize $80$} 
\psfrag{ 100l}[r][r]{\footnotesize $100$} 
\psfrag{ 1}    [c][c]{\footnotesize $1$} 
\psfrag{ 1.01} [c][c]{\footnotesize $1.01$} 
\psfrag{ 1.02} [c][c]{\footnotesize $1.02$} 
\psfrag{ 1.03} [c][c]{\footnotesize $1.03$} 
\psfrag{ 1.04} [c][c]{\footnotesize $1.04$} 
\psfrag{ 1.06} [c][c]{\footnotesize $1.06$} 
\psfrag{ 1.08} [c][c]{\footnotesize $1.08$} 
\psfrag{ 1.1}  [c][c]{\footnotesize $1.1$} 
\psfrag{f} [c][c]{\footnotesize Tip force $f$} 
\psfrag{l} [l][l]{\footnotesize $\bar\lambda$} 
\psfrag{a/l0.05} [l][l]{\scriptsize $a_0/l_0=0.05$} 
\psfrag{a/l0.1}  [l][l]{\scriptsize $a_0/l_0=0.1$} 
\psfrag{a/l0.15} [l][l]{\scriptsize $a_0/l_0=0.15$} 
\psfrag{a/l0.2}  [l][l]{\scriptsize $a_0/l_0=0.2$} 
\psfrag{B}   [l][l]{\scriptsize $B=1$} 
\psfrag{EA}  [l][l]{\scriptsize $EA=1$} 
\psfrag{B/EA1}   [l][l]{\scriptsize $B/EA=1$} 
\psfrag{B/EA2}   [l][l]{\scriptsize $B/EA=2$} 
\psfrag{B/EA3}   [l][l]{\scriptsize $B/EA=3$} 
\psfrag{B/EA4}   [l][l]{\scriptsize $B/EA=4$} 
\psfrag{a} [l][l]{\normalsize $a$)} 
\psfrag{b} [l][l]{\normalsize $b$)} 
\centering
\includegraphics[width=14cm]{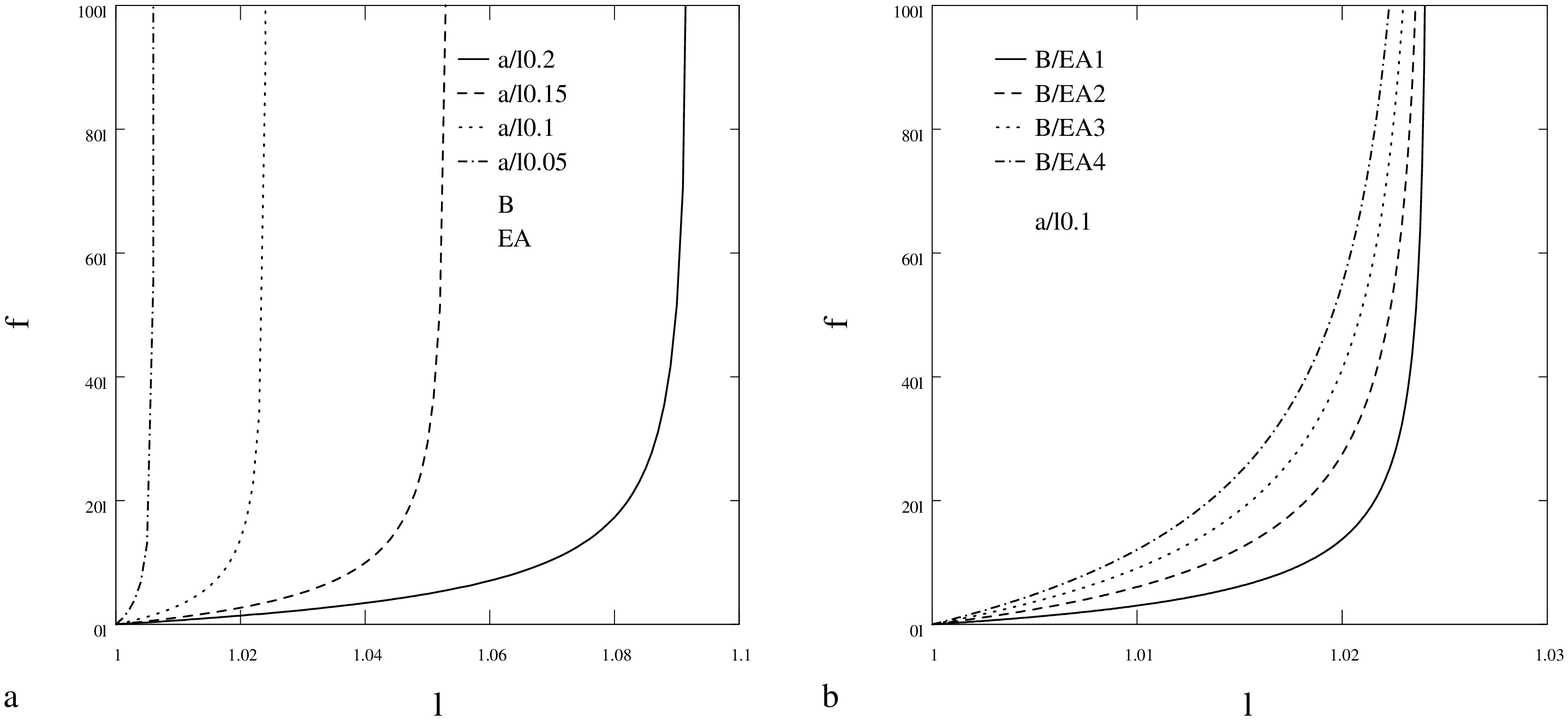}
\caption{Sinusoidal inextensible elastica.  Comparison of the
  $f{-}\bar\lambda$ curves for globally inextensible sinusoidal
  elasticas having different $a$) $a_0/l_0$ and $b$) $B/EA$ ratios.  
}
\label{fig5}
\end{figure}

First, we consider a sinusoidal elastica with the additional
global inextensibility assumption given in
(\ref{gloinextensibility}). In Figure \ref{fig5}a the influence of the 
ratio $a_0/l_0 \in [0.05,0.2]$  on the $f{-}\bar\lambda$ curve is
depicted while keeping the material parameters fixed at $B=1$ and $EA=1$. This
ratio proves crucial in determining the value of 
stretch at which the heel occurs. The higher the ratio $a_0/l_0$, the
longer the toe region preceding the heel. In other words, this
parameter determines the value of $\bar{\lambda}$ where the influence
of the bending mechanism starts to 
diminish and the axial extension begins to govern the
$f{-}\bar\lambda$ curve. In
order to demonstrate the sensitivity of the $f{-}\bar\lambda$ curve to the
bending stiffness, the ratio of bending stiffness to axial stiffness,
$B/EA$, is varied from $1$ to $4$ (Figure \ref{fig5}b). An increase
in the ratio $B/EA$ scales the curve's ordinates ($f$-values), and
therefore the 
transition in the heel region becomes more gradual. However, the value
of the locking stretch is not influenced by the changes in the ratio $B/EA$.

\begin{figure}[ht]
\psfrag{ 0l}  [r][r]{\footnotesize  $0$} 
\psfrag{ 100l}[r][r]{\footnotesize $100$} 
\psfrag{ 200l}[r][r]{\footnotesize $200$} 
\psfrag{ 300l}[r][r]{\footnotesize $300$} 
\psfrag{ 400l}[r][r]{\footnotesize $400$} 
\psfrag{ 500l}[r][r]{\footnotesize $500$} 
\psfrag{ 1}    [c][c]{\footnotesize $1$} 
\psfrag{ 1.5}  [c][c]{\footnotesize $1.5$} 
\psfrag{ 2}    [c][c]{\footnotesize $2$} 
\psfrag{ 2.5}  [c][c]{\footnotesize $2.5$} 
\psfrag{f} [c][c]{\footnotesize Tip force $f$} 
\psfrag{l} [l][l]{\footnotesize $\bar\lambda$} 
\psfrag{a/l0.02} [l][l]{\scriptsize $a_0/l_0=0.02$} 
\psfrag{a/l0.08} [l][l]{\scriptsize $a_0/l_0=0.08$} 
\psfrag{a/l0.14} [l][l]{\scriptsize $a_0/l_0=0.14$} 
\psfrag{a/l0.2}  [l][l]{\scriptsize $a_0/l_0=0.2$} 
\psfrag{a/l0.1}  [l][l]{\scriptsize $a_0/l_0=0.1$} 
\psfrag{EA/B30}  [l][l]{\scriptsize $EA/B=30$} 
\psfrag{EA/B120}  [l][l]{\scriptsize $EA/B=120$} 
\psfrag{EA/B210}  [l][l]{\scriptsize $EA/B=210$} 
\psfrag{EA/B300}  [l][l]{\scriptsize $EA/B=300$} 
\psfrag{a} [l][l]{\normalsize $a$)} 
\psfrag{b} [l][l]{\normalsize $b$)} 
\centering
\includegraphics[width=14cm]{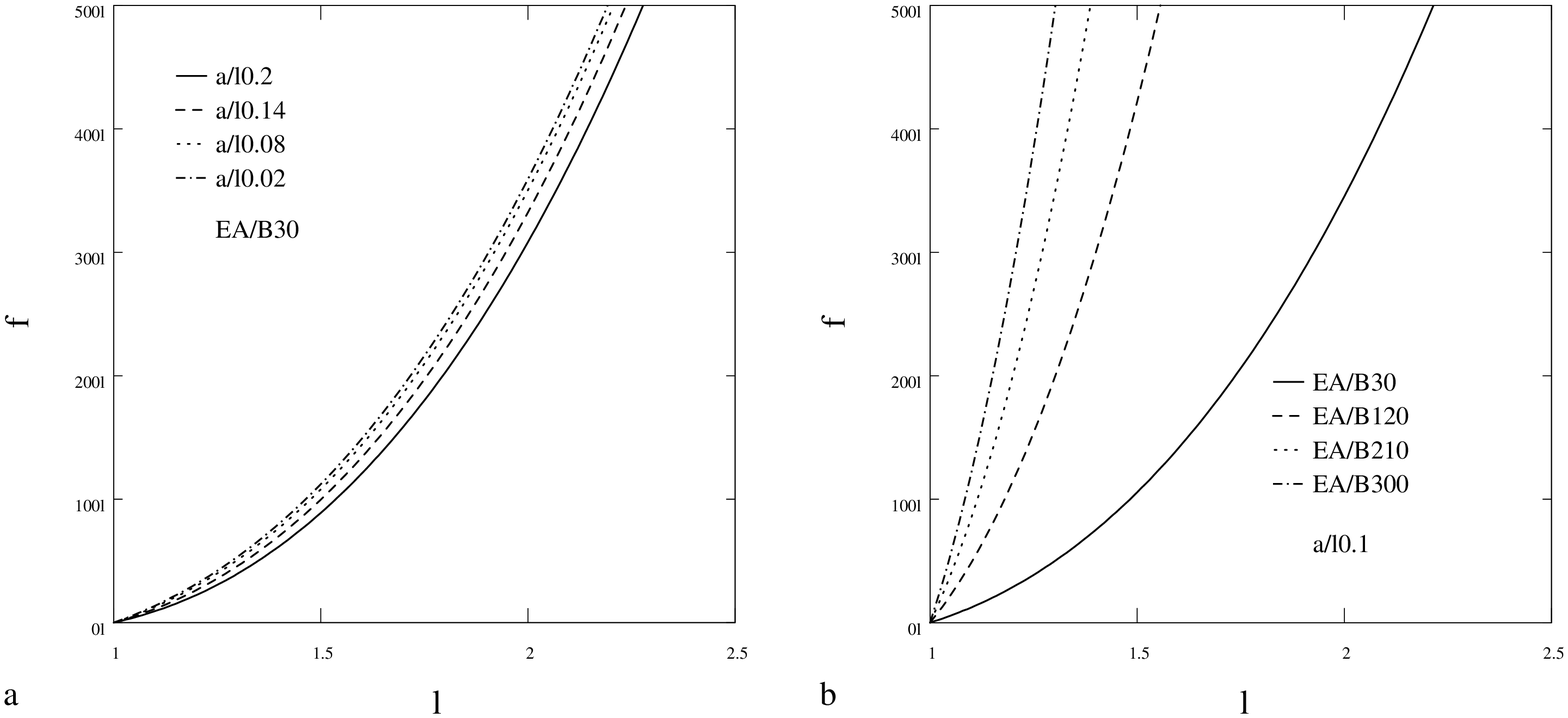}
\caption{Sinusoidal elastica surrounded by a planar incompressible
 medium. Comparison of the $f{-}\bar\lambda$ curves for  different
 $a$) $a_0/l_0$ and $b$) $EA/B$ ratios.
}
\label{fig6}
\end{figure}

\begin{figure}[ht]
\psfrag{ 0l}  [r][r]{\footnotesize  $0$} 
\psfrag{ 100l}[r][r]{\footnotesize $100$} 
\psfrag{ 200l}[r][r]{\footnotesize $200$} 
\psfrag{ 300l}[r][r]{\footnotesize $300$} 
\psfrag{ 400l}[r][r]{\footnotesize $400$} 
\psfrag{ 500l}[r][r]{\footnotesize $500$} 
\psfrag{ 1}    [c][c]{\footnotesize $1$} 
\psfrag{ 1.5}  [c][c]{\footnotesize $1.5$} 
\psfrag{ 2}    [c][c]{\footnotesize $2$} 
\psfrag{ 2.5}  [c][c]{\footnotesize $2.5$} 
\psfrag{f} [c][c]{\footnotesize Tip force $f$} 
\psfrag{l} [l][l]{\footnotesize $\bar\lambda$} 
\psfrag{a/l0.02} [l][l]{\scriptsize $a_0/l_0=0.02$} 
\psfrag{a/l0.08} [l][l]{\scriptsize $a_0/l_0=0.08$} 
\psfrag{a/l0.14} [l][l]{\scriptsize $a_0/l_0=0.14$} 
\psfrag{a/l0.2}  [l][l]{\scriptsize $a_0/l_0=0.2$} 
\psfrag{a/l0.1}  [l][l]{\scriptsize $a_0/l_0=0.1$} 
\psfrag{EA/B30}  [l][l]{\scriptsize $EA/B=30$} 
\psfrag{EA/B120}  [l][l]{\scriptsize $EA/B=120$} 
\psfrag{EA/B210}  [l][l]{\scriptsize $EA/B=210$} 
\psfrag{EA/B300}  [l][l]{\scriptsize $EA/B=300$} 
\psfrag{a} [l][l]{\normalsize $a$)} 
\psfrag{b} [l][l]{\normalsize $b$)} 
\centering
\includegraphics[width=14cm]{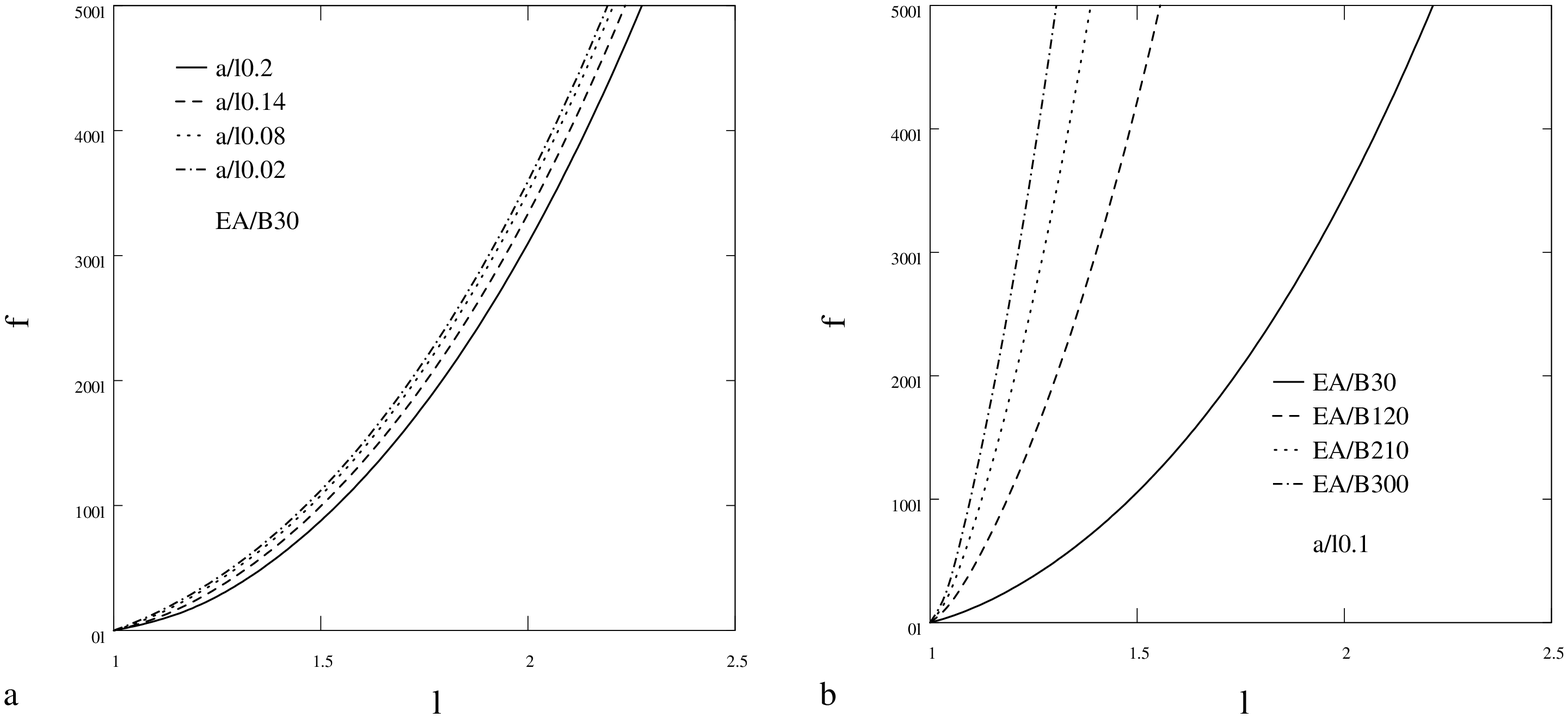}
\caption{Sinusoidal elastica deforming by attaining a stationary strain
 energy state. Comparison of the $f{-}\bar\lambda$ curves for 
 different $a$) $a_0/l_0$ and $b$) $EA/B$ ratios.  
}
\label{fig7}
\end{figure}

In the last two cases we consider the planar incompressible and stationary energy
sinusoidal elasticas. Figures \ref{fig6}a and \ref{fig7}a
present the influence of the change in ratio $a_0/l_0$ on 
the $f{-}\bar\lambda$ curves of the respective cases. Like the 
inextensible case the ratio $a_0/l_0$ is varied within
the interval $[0.05,0.2]$ while the value of the ratio
$EA/B$ is kept fixed at $30$. Clearly, the
$f{-}\bar\lambda$ curves for the 
planar incompressible and stationary energy cases do not exhibit a sharp
transition to
stiffening behavior. This is in contrast with the inextensible
case in Figure \ref{fig5}. Variation of the ratio $a_0/l_0$ does not
cause significant change in the shape of the curves. 

The sensitivity of the $f{-}\bar\lambda$ curves for the separate cases
to changes in material parameters $EA$ and $B$ is presented in Figures 
\ref{fig6}b and \ref{fig7}b, respectively. The ratio
$EA/B$ varies in the range $[30,300]$. The axial
stiffening is clearly reflected in the curves. No striking
shape change is observed. We draw attention to the fact that the 
$f{-}\bar\lambda$ curves in Figures \ref{fig6} and \ref{fig7} for the
planar incompressible and stationary energy cases of the sinusoidal elastica 
are quite similar. The reasons for this similarity have
been already outlined in Remark 8.

\subsection{Comparison with experiment}
\begin{figure}[thb]
\psfrag{ 0l}    [r][r]{\footnotesize  $0$} 
\psfrag{ 0.2l}  [r][r]{\footnotesize  $0.2$} 
\psfrag{ 0.4l}  [r][r]{\footnotesize  $0.4$} 
\psfrag{ 0.6l}  [r][r]{\footnotesize  $0.6$} 
\psfrag{ 0.8l}  [r][r]{\footnotesize  $0.8$} 
\psfrag{ 1l}    [r][r]{\footnotesize  $1$} 
\psfrag{ 1.2l}  [r][r]{\footnotesize  $1.2$} 
\psfrag{ 1.4l}  [r][r]{\footnotesize  $1.4$} 
\psfrag{ 1}    [c][c]{\footnotesize $1$} 
\psfrag{ 1.1}  [c][c]{\footnotesize $1.1$} 
\psfrag{ 1.2}  [c][c]{\footnotesize $1.2$} 
\psfrag{ 1.4}  [c][c]{\footnotesize $1.4$} 
\psfrag{ 1.6}  [c][c]{\footnotesize $1.6$} 
\psfrag{ 1.8}  [c][c]{\footnotesize $1.8$} 
\psfrag{ 2}    [c][c]{\footnotesize $2$} 
\psfrag{f} [c][c]{\footnotesize Nominal Stress $\:[MPa]$} 
\psfrag{l} [l][l]{\footnotesize $\bar\lambda\:[-]$} 
\psfrag{exp} [l][l]{\scriptsize Experiment}  
\psfrag{sin_ie} [l][l]{\scriptsize Sinusoidal}  
\psfrag{cir_me} [l][l]{\scriptsize Circular-Arc}  
\centering
\includegraphics[width=7cm]{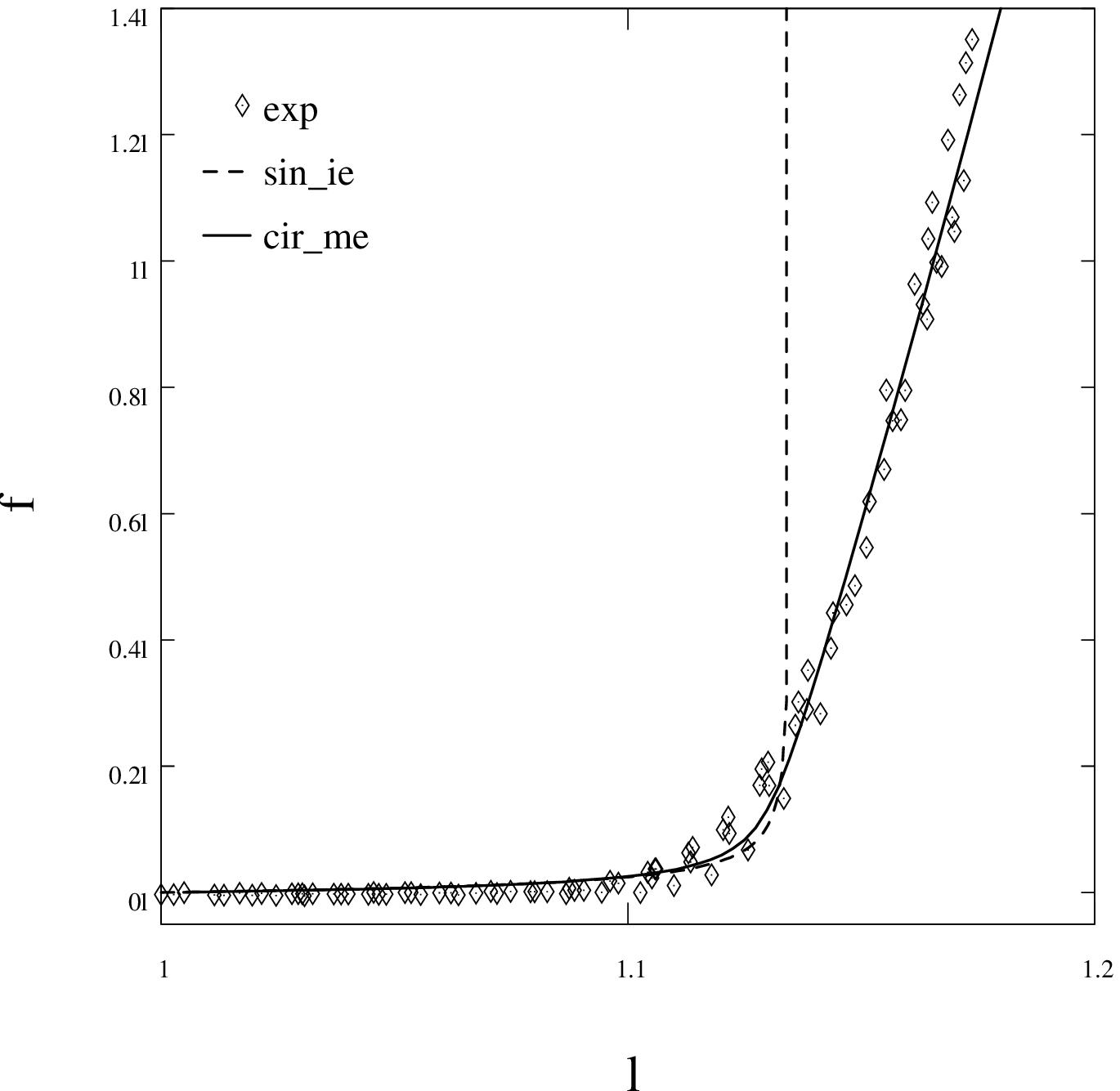}
\caption{Simulations of experimental data by the inextensible
  sinusoidal elastica ($a_0/l_0=0.245\;, B/EA=25\:\mathrm{mm}^{-1}$) and the
  circular-arc elastica attaining a stationary strain energy state
  ($R=0.013\:\mathrm{mm}\;,\theta_{0}=4\pi/ 15\;,
  EA/B=7\times 10^{6}\:\mathrm{mm}^{-2}$). In the elastica models $R,
  a_0$ and $l_0$ were in mm
  $B$ in $\mathrm{N}/\mathrm{mm}$ and $EA$ in $\mathrm{MPa}/\mathrm{mm}$.  
}
\label{fig8}
\end{figure}

In the preceding section the sensitivities of the $f-\bar{\lambda}$
curves to geometric and material parameters have been discussed for
both the circular-arc and the sinusoidal elasticas subjected to two
additional kinematic assumptions, and the stationary strain energy
assumption. In this section we carry out a comparison with 
data reported by \citet{freed+doehring05} (Figure \ref{fig8}). 
These data correspond to uniaxial extension experiments on five
chordae tendineae from procine mitral valves. They demonstrate a long
toe region relative to the maximum stretch in each experiment. At
the heel $(\bar\lambda_\mathrm{heel}\approx 1.13)$,  the nominal
stress--stretch curve stiffens sharply to a larger
slope. From the results in Figures \ref{fig0}--\ref{fig7} of the
preceding parameter study, it is apparent that this behavior can only
be captured either by the inextensible sinusoidal elastica, or the
circular-arc elastica attaining a stationary energy state. Figure
\ref{fig8} compares the experiment with these two models with the
material parameters given in the caption. Both the sinusoidal and the
circular-arc models successfully match the data in the toe region. The
inextensible sinusoidal model can also predict the upturning region,
but its stiffness beyond the heel region rapidly diverges and fails to
match the experimental results. In the case of the circular-arc model,
the value of 
$\theta_{0}$ can be analytically determined from the
macro-stretch value at the heel. We solve for $\theta_0$ such that
$\theta_0/\sin\theta_{0} = \bar\lambda_\mathrm{heel}=1.13$.
This gives the initial angle $\theta_{0}\approx 4\pi/ 15$, and
the ratio of the axial and bending stiffness $EA/B$ can be tuned to
match the slopes of the regions just preceding and succeeding
the heel region. The initial radius $R$ is varied to match the
sharpness of the slope change at the heel region. The 
comparison of the stationary strain energy circular-arc elastica and the
experimental data clearly illustrates that the proposed model
quantitatively captures the experimental data with just a few
parameters: $\theta_0, R, B$ and $EA$, all of which are very
well-motivated physically. Clearly, other such experimental data can be matched
without difficulty.

\section{Macroscopic material model incorporating the elastica}
\label{sect3}

\subsection{Continuum strain energy density function at the macroscale}
The contribution to the overall strain energy density function due to
the collagen fibrils embedded in a nearly 
incompressible viscous medium is obtained by summing up the free
energies of individual elastica-like fibrils,
\begin{equation}
\Psi_\mathrm{col}= \frac{N}{A_0\,l_0}\widetilde W(g)\:.
\label{Psiclg}
\end{equation}
\noindent With the unit vector $\be$ denoting the average orientation of 
collagen fibrils, the macroscopic stretch in this direction is
obtained by  $\bar\lambda=|\bF\be|$, where $\bF$ is the deformation
gradient tensor. In the context of anisotropic
elasticity, especially transverse isotropy, it is common to define
structural tensors $\bM:=\be \otimes\be$ for the construction of strain
energy density functions formulated in terms of additional invariants. The
derivatives of these invariants are then used as tensor generators in
the stress response functions. In the present case,
$I_4:=\bC:\bM=\bar\lambda^2$ is the relevant invariant ($\bC$ being
the right Cauchy-Green tensor). We continue to use an
affine relation between the macro--stretch and tip displacement of fibrils,
i.e. $\bar\lambda=1 + g/l_0\,$, where $l_0$ is the half wavelength of
a sinusoidal elastica, and $l_0 = 2R\sin\theta_0$ for a circular-arc
elastica. The cross-sectional area of the tissue that contains $N$
such fibrils is $A_0$. With this relation in hand, the
contribution to the total second Piola--Kirchhoff stress tensor due to
the stretching of collagen fibrils,
$\bS^\mathrm{col}=2\partial\Psi_\mathrm{col}/\partial\bC$, can be obtained as 
\begin{equation}
\bS^\mathrm{col} = N\,\frac{f(g)/\bar\lambda}{A_0}\; \bM
\label{PK2clg}
\end{equation}
where the results $\partial g/\partial\bar\lambda=l_0\;,$ 
$2\partial\bar\lambda/\partial I_4 =1/\bar\lambda\;,$
$\partial I_4/\partial\bC=\bM$ and the definition 
$f(g):=\partial\widetilde W /\partial g$ have been used. Then, the
nominal stress tensor $\bP^\mathrm{col}$ readily follows from $\bP^\mathrm{col}=\bF\bS^\mathrm{col}$:
\begin{equation}
\bP^\mathrm{col} = N\,\frac{f(g)}{A_0}\; \tilde{\be} \otimes\be
\label{PK1clg}
\end{equation}
where $\bF\be=\bar\lambda\tilde{\be}$ and $|\tilde{\be}|=1\,.$ 

We now turn our attention to the convexity of the strain energy density
$\Psi_\mathrm{col}=\hat\Psi_\mathrm{col}(I_4)$, in order to have a
basic understanding of its stability properties. The convexity condition demands the
positive definiteness of the first elasticity tensor $\mathbb{A}^\mathrm{col}$
\begin{equation}
\bH:\mathbb{A}^\mathrm{col}:\bH\ge0
\qquad 
\forall\bH\in\mathbb{M}^{3\times 3} 
\quad\mathrm{and}\quad
\mathbb{A}^\mathrm{col}:=\frac{\partial^2\hat\Psi_\mathrm{col}}{\partial\bF\partial\bF}\:,
\label{convexity}
\end{equation}
\noindent where $\mathbb{M}^{3\times 3}$ is the space of second-order
tensors in $\mathbb{R}^3$.

The explicit form of the first elasticity tensor $\mathbb{A}^\mathrm{col}$
can be obtained by the chain rule as 
\begin{equation}
\mathbb{A}^\mathrm{col}:=
\hat\Psi'_\mathrm{col}\:\frac{\partial^2I_4}{\partial\bF\partial\bF}  
+ \hat\Psi''_\mathrm{col}\:
\frac{\partial I_4}{\partial\bF}\otimes\frac{\partial I_4}{\partial\bF}
\label{moduli}
\end{equation}
where the superscript $(\cdot)'$ denotes the derivatives with respect
to $I_4$. The quadratic product  
of $\mathbb{A}^\mathrm{col}$ with $\bH$ involves the terms   
\begin{equation}
\bH:\frac{\partial^2I_4}{\partial\bF\partial\bF}:\bH=  
2\Vert\bH\be\Vert^2\ge 0\;,
\qquad
\left(\frac{\partial I_4}{\partial\bF}:\bH\right)^2 =
(2\bH\be\cdot\bF\be)^2 \ge 0. 
\label{contractions}
\end{equation}
Note that both terms are non-negative. The local convexity condition 
(\ref{convexity}) of the free energy $\hat\Psi_\mathrm{col}$ reduces to
\begin{equation}
2\hat\Psi'_\mathrm{col}\:\Vert\bH\be\Vert^2
+ 4\hat\Psi''_\mathrm{col}\:(\bH\be\cdot\bF\be)^2
\ge 0 \:.
\label{convexity2}
\end{equation}
Based on (\ref{contractions}),
non-negativity of both $\hat\Psi'_\mathrm{col}$ and
$\hat\Psi''_\mathrm{col}$  is sufficient to
fulfill the convexity condition (\ref{convexity2}), though not necessary. The 
explicit forms of the derivatives are 
$\hat\Psi'_\mathrm{col}=Nf(g)/(2A_0\bar\lambda)$ and 
$\hat\Psi''_\mathrm{col}=
N l_0 (f'(g)l_0+ gf'(g) -f(g))/(4A_0\bar\lambda^2(g+l_0))$. If we
assume that collagen fibrils can carry only tensile loads, the
positiveness of $\hat\Psi'_\mathrm{col}$ is satisfied  
identically for $f(g)\ge 0\;.$ Furthermore, the convexity of
$\hat\Psi_\mathrm{col}$ with respect to $g$ ensures that $f'(g)\ge 0\;.$
Thus, it is now sufficient to show that the term $gf'(g) -f(g)\ge 0$
in $\hat\Psi''_\mathrm{col}\;.$ This condition can be 
obtained starting from the convexity condition for $f(g)$ with respect
to $g$, i.e. $f''(g)\ge 0\;.$ For positive values of $g$, we have
$gf''(g)\ge 0\;.$ Integration of $gf''(g)\ge 0$ by parts yields
$\int_0^g gf''(g)=gf'(g)|_0^g-\int_0^g f'(g)\ge 0\;.$ For $f(0)=0$, we
obtain the sought form $gf'(g)-f(g)\ge 0\;.$ Therefore, convexity of
both the $\widetilde W$ and $f(g)=\partial\widetilde W /\partial g$
guarantees the local convexity of the macroscopic free energy function 
$\hat\Psi_\mathrm{col}\;.$

When $\hat\Psi_\mathrm{col}$ is combined by rule-of-mixtures with the
strain energy density function of the surrounding matrix medium, the
above results completely characterize the influence upon the convexity
of the overall composite, leaving open only the question of convexity
of the matrix material. However, as pointed out by a reviewer, the actual
interaction between the collagen fibrils and matrix involves shearing
of the matrix and consideration of the rate of decay of shear fields
with distance from a loaded fibril. There is then the possibility of
more complex interaction between strain energy of the matrix and of the
elastica-like fibrils. The convexity arguments presented here will
then have to be refined by further considerations of fibril-matrix
interactions. 
\section{Closing remarks}
\label{closing}

The primary aim of this paper is a discussion of the characteristic
soft tissue response in the context
of the elastica-like mechanical behavior of slender fibrillar
structures in these 
tissues. The models are applicable to tendons, skin and the passive
response of muscle. While entropic elasticity-based models can also model
this characteristic soft tissue response---especially the locking
behavior---there are strong physical and physiological reasons 
to surmise that this is the wrong approach to adopt. A direct solution
of the shape and force in a 
deforming elastica requires the solution of a (``highly'') nonlinear,
fourth-order partial differential equation. The simplification
used here is that judiciously-chosen additional assumptions on the
kinematics and on the energy state can lead to force-deformation
response functions for the 
elastica. This is the central thesis advanced in this paper. Beyond
this, the paper is concerned with an enumeration of 
two families of shapes (circular arcs and sinusoidal half-periods) of
the deforming elastica, and three possible 
additional assumptions: inextensibility,
macroscopic planar incompressibility, and stationarity of strain
energy. The motivations for each of these additional assumptions are
well-founded in a physical sense. Their suitability in matching a set
of experimental force-deformation curves has been examined. On the
basis of the current limitation to elastic effects, it emerges
that the elastica deforming as a circular arc, and maintaining itself
in a state of stationary strain energy in each configuration (parametrized by
overall elongation) can resolve the experimental data to a high
degree of precision.\footnote{We have not demonstrated
  quantitative error measures since no significant physical insight is
  gained by doing so.} The parameters used are the two stiffnesses---bending
and axial, and two geometric parameters that determine the shape of
the undeformed elastica. These can be easily determined from
mechanical experiments and micrographs, and compared with the values
obtained for the best fit. Such an exercise would be a strong
validation of these models. We note that the circular-arc elastica
with stationary energy matches the experimental data very well in
Figure \ref{fig8} with initial radius $R = 0.013\:\mathrm{mm}$
($13\:\mu\mathrm{m}$), and
initial central angle $2\theta_0 = 8\pi/15$ ($96^\circ$). This gives a
wavelength of $4R\sin\theta_0 = 38.64\;\mu\mathrm{m}$ which seems
very reasonable, given that collagen fibrils are typically found to
have wavelengths between $10$ and $50\:\mu\mathrm{m}$ as in
\citet{Screenetal:2004} and \citet{ProvenzanoVanderby:2006}. The
importance of matrix shear lag and its influence on convexity,
inelastic effects such as the viscous friction as 
collagen fibrils move relative to the surrounding proteoglycans,
viscoelasticity of the collagen fibrils themselves and proteoglycans,
and slippage of 
fibrils under larger forces, must not be overlooked, however.

It should also be quite clear, that the development here has complete
relevance for many types of slender filamentous structures, from
carbon nanotubes, 
through underwater cables to oil pipelines. The class of continuum
strain energy density 
functions so developed in Section \ref{sect3} is applicable to any
composite consisting of mainly unidirectional, elastica-like
reinforcing fibers in a matrix.   

\bibliography{mybib}
\bibliographystyle{elsart-harv}
\end{document}